\documentclass[journal]{IEEEtran}
\usepackage{booktabs}
\usepackage{subfig}
\usepackage{float}
\usepackage{array} 
\usepackage{amssymb}
\usepackage{graphicx}
\usepackage{float}
\usepackage{amsmath}
\usepackage{amsthm}
\usepackage{cite}

\newtheorem{theorem}{Theorem}
\newtheorem{lemma}{Lemma}
\newtheorem{proposition}{Proposition}
\newtheorem{conjecture}{Conjecture}
\ifCLASSINFOpdf
\else
\fi
\begin{document}
%
\title{Performance Analysis of Cooperative Integrated Sensing and Communications for 6G Networks\\ }
%
%
%

\author{Dongsheng Sui, Cunhua Pan, Hong Ren, Jiahua Wan, Liuchang Zhuo,~\IEEEmembership{Member,~IEEE,} Jing Jin, Qixing Wang and Jiangzhou Wang,~\IEEEmembership{Fellow,~IEEE}
\thanks{D. Sui is with the Bell Honers School, Nanjing University of Posts and Telecommunications, Nanjing, Jiangsu 210003, China. (e-mail: b21041031@njupt.edu.cn)}
\thanks{C. Pan, H. Ren, J. Wan and L. Zhuo are with the National Mobile Communications Laboratory, Southeast University, Nanjing, Jiangsu 211189, China. (e-mail: cpan, hren, wanjiahua, 230248854@seu.edu.cn)}
\thanks{J. Jin and Q. Wang are with the Future Research Laboratory, China Mobile Research Institute, Beijing 100053, China. (e-mail: jinjing@chinamobile.com; wangqixing@chinamobile.com)}
\thanks{J. Wang is with the National Mobile Communications Laboratory, Southeast University, and Purple Mountain Laboratories, Nanjing, Jiangsu 211119, China.}
}

\maketitle

\begin{abstract}
In this work, we aim to effectively characterize the performance of cooperative integrated sensing and communication (ISAC) networks and to reveal how performance metrics relate to network parameters. To this end, we introduce a generalized stochastic geometry framework to model the cooperative ISAC networks, which approximates the spatial randomness of the network deployment. Based on this framework, we derive analytical expressions for key performance metrics in both communication and sensing domains, with a particular focus on communication coverage probability and radar information rate. The analytical expressions derived explicitly highlight how performance metrics depend on network parameters, thereby offering valuable insights into the deployment and design of cooperative ISAC networks. In the end, we validate the theoretical performance analysis through Monte Carlo simulation results. Our results demonstrate that increasing the number of cooperative base stations (BSs) significantly improves both metrics, while increasing the BS deployment density has a limited impact on communication coverage probability but substantially enhances the radar information rate. Additionally, increasing the number of transmit antennas is effective when the total number of transmit antennas is relatively small. The incremental performance gain reduces with the increase of the number of transmit antennas, suggesting that indiscriminately increasing antennas is not an efficient strategy to improve the performance of the system in cooperative ISAC networks.  
\end{abstract}


%
\IEEEpeerreviewmaketitle

\section{Introduction}
\IEEEPARstart{W}{ith} the advancement of researches on sixth-generation (6G), a consensus has emerged in academia and industry: sensing capabilities are evolving into critical requirements for future wireless systems due to their extensive applications in target positioning, velocity measurement, and imaging \cite{8869705,9705498}. To address these demands, integrated sensing and communication (ISAC) has been identified as a key enabler for next-generation wireless networks, enabling simultaneous target sensing with sub-meter-level accuracy and user-centric communications with high throughput and ultra-low latency \cite{9945983}. Moreover, the development of communication and sensing technologies has historically progressed independently, with limited cross-domain interaction. As communication frequencies increasingly shift toward higher bands, ISAC technology has quickly emerged as a research focus due to its unique hardware resource-sharing architecture. This architecture facilitates dual utilization of radar spectrum resources and infrastructure components, thereby enhancing both spectral and hardware efficiency \cite{9557830}. ISAC supports diverse applications, including intelligent transportation, smart cities, smart manufacturing \cite{9830717,9755276}, and more broadly, the Internet of Things (IoT) \cite{7465731}. However, current investigations remain largely confined to single-node ISAC architectures \cite{10566041,10054402,10527368,10644091,9585321,10012421}, with only limited attention dedicated to cooperative ISAC architectures. \par
In contrast to single-node ISAC architectures, which require costly hardware upgrades to support full-duplex operation and face challenges from stringent self-interference cancellation requirements \cite{9540344,10159012}, cooperative ISAC architectures leverage collaborative signal processing to mitigate interference, thereby improving system performance \cite{9860459}. Additionally, single-node ISAC architectures often suffer from limited robustness in target sensing and constrained throughput in user communication, consequently failing to meet the stringent demands for 6G application scenarios \cite{10892128,10540489}. Cooperative ISAC architectures achieve high-throughput, ultra-reliable, and low-latency communication by utilizing coordinated multi-point (CoMP) transmission/reception techniques for efficient interference management. In addition, they enable ultra-precise, high-resolution, and robust sensing capabilities through multi-perspective sensing data fusion and reinforced acquisition of effective echo signals \cite{10833779,ren2024optimal}. However, cooperative ISAC also introduces new challenges associated with system deployment and resource allocation. Moreover, increased signaling overhead from the close coordination among base stations (BSs) \cite{10726912} and  restricted backhaul capacity \cite{10680299} also constrain its development. Therefore, it is critical to comprehensively investigate both the communication and sensing performance, as well as the impacts of system parameters on the performance metrics, to achieve a balance between performance gains and operational costs. This motivates research on cooperative ISAC design to systematically  quantify and improve collaborative performance gains, establishing a foundational framework for advancing cooperative ISAC network architectures.\par
However, accurately quantifying the system performance and effectively assessing the impacts of system parameters on cooperative ISAC pose significant challenges due to the inherent spatial randomness in BSs' distribution, communication users' locations, and sensing targets' positions. Traditional performance analysis methods struggle to directly and explicitly evaluate such systems due to deployment randomness, relying instead on complex time-consuming simulations that lack analytical insights to guide performance optimization and system design. Stochastic geometry has emerged as a powerful mathematical framework for modeling spatial distributions and deriving quantifiable system metrics, and it has been widely applied in communication-only systems \cite{6287527,6042301,6658810}. For example, \cite{6596082} demonstrated coverage probability analysis and rate comparisons in downlink multi-antenna dense heterogeneous networks using stochastic geometry. \cite{6881662} studied load balancing in multi-antenna heterogeneous networks through SINR-maximizing cell selection rules and analyzed its impact on network coverage probability and user rate. Additionally, radar system studies have also extensively utilized stochastic geometry to model the spatial distributions of objects and radar installations \cite{9542942,8587144}. For instance, \cite{9580712} derived analytical expressions for optimal radar bandwidth, transmit power, and detection thresholds under both noise-limited and clutter-limited conditions by using the Swerling-1 model with the generalized Weibull distribution to unify both exponential and Rayleigh cases. To overcome the insufficiency of detectable positioning signals, \cite{7412750} employed stochastic geometry to derive precise analytical expressions for unique localization probabilities in collaborative/non-collaborative scenarios with varying shadowing and frequency reuse conditions. A common approach of these studies involves modeling practical system elements through a spatial point process, and then applying stochastic geometry tools like Laplace transforms and probability generating functional (PGFL) for theoretical performance analysis. With the continued integration of sensing and communication, stochastic geometry is poised to become indispensable for addressing cooperative ISAC system challenges through rigorous spatial modeling and analytical derivations.\par 
Recent studies have applied stochastic geometry to analyze cooperative ISAC networks. For instance, \cite{10556618} investigated coverage and ergodic rate in a single-site communication and multi-site sensing ISAC framework. \cite{10769538} derived a scaling law for the Cramér-Rao lower bound (CRLB) of localization accuracy and analyzed communication rates. \cite{10735119} proposed a coordinated beamforming strategy to mitigate interference in ISAC networks, providing valuable insights into resource allocation. However, the work in \cite{10556618} assumed sensing targets as a user equipped with an active antenna, which differed from passive targets in practical scenarios. Moreover, it failed to address communication coverage probability analysis in the cooperative settings. More importantly, existing literature has largely overlooked enhancing communication coverage capacity through close coordination among BSs to optimize ISAC networks performance. Communication coverage probability, a key performance metric that reflects interference resistance capability and reliable connectivity maintenance with BSs, plays a pivotal role in guiding the optimal cooperation approach in interference-limited scenarios. Furthermore, radar information rate, defined as the mutual information between target impulse response and received measurements, effectively quantifies radar detection capabilities. A higher radar information rate indicates higher accuracy in estimating target parameters \cite{rir,rir2}. Compared to traditional metrics like CRLB, radar information rate offers advantages in analytical tractability and computational simplicity. Notably, the CRLB-based analysis for target detection in \cite{10769538} focused exclusively on the accuracy of the estimated position, while failing to confirm whether a target actually exists at the estimated location. In contrast, combining higher radar information rates with lower CRLB values not only confirms the existence of a target that requires estimation but also ensures the accuracy of the target’s estimated position. Thus, adopting radar information rate as a sensing performance metric is crucial for comprehensively assessing system detection capabilities.\par  
Building on the preceding discussions, we propose a generalized stochastic geometry framework to model cooperative ISAC networks that integrate the CoMP transmission for cooperative communication with multi-static radar configurations for cooperative sensing. Through our framework, we quantify ISAC performance using two key metrics: communication coverage probability and radar information rate. Stochastic geometry and probability theory are employed to examine the critical impact of system parameters on performance optimization in ISAC networks. By leveraging stochastic geometry, we rigorously model the inherent spatial randomness in cooperative ISAC systems, particularly the spatial distribution of BSs. Using derived analytical expressions for the aforementioned metrics, we conduct comprehensive performance analysis to reveal how BS deployment density and the number of cooperative BSs affect both metrics. Furthermore, compared with prior works on single-node ISAC \cite{10695929,10683162}, this work quantitatively demonstrates significant performance gains achieved through cooperative strategies. The main contributions of this paper are summarized as follows.
\begin{itemize}
    \item \textbf{Generalized stochastic geometry framework}: We develop a stochastic geometry-based analytical framework to model and evaluate the performance of cooperative ISAC networks. The proposed framework explicitly captures the inherent spatial randomness in practical systems and facilitates tractable analysis of both communication and sensing metrics under cooperative operation. 
    \item \textbf{Coverage probability and radar information rate derivations}: We propose a small-scale fading approximation method to derive an analytical expression for the communication coverage probability in multi-BS cooperative communication networks. We further derive a closed-form expression for the coverage probability under specific scenarios, which is shown to be independent of the BS deployment density. Additionally, the Laplace transforms of the desired echo signal power and the interference power are rigorously derived using moment-generating functions and the statistical distribution of distances between interfering BSs and the echo-receiving BS. Based on these results, a tractable analytical expression for the radar information rate is subsequently obtained.
    \item \textbf{Insights into BS deployment strategy}: Our results demonstrate that the BS deployment density exerts minimal impact on the communication coverage probability while substantially improving the radar information rate. This indicates that simply increasing BS deployment density offers limited benefits for enhancing communication coverage probability in cooperative ISAC networks, despite its effectiveness in boosting radar sensing performance. In contrast, the number of cooperative BSs exhibits comparable positive effects on both metrics. Specifically, increasing the number of cooperative BSs engaged in user communication or target sensing leads to simultaneous improvements in coverage probability and radar information rate, reflecting greater system robustness against interference and enhanced sensing precision. Furthermore, increasing the number of transmit antennas is effective for enhancing cooperative ISAC performance when the total number of transmit antennas is relatively small. However, this strategy becomes less efficient as marginal performance gains diminishes with larger number of antennas.
\end{itemize}\par
The structure of this paper is organized as follows. Section II presents the cooperative ISAC system model and defines the key performance metrics. Section III develops analytical approximations for these metrics based on stochastic geometry and probability theory, and derives solutions for a special case. Section IV validates the theoretical accuracy of the derived results through comparisons with Monte Carlo simulations and offers practical insights into BS deployment strategies. Section V concludes the paper by summarizing key findings.\par
Notations: In this paper, bold lowercase letters denote vectors and bold uppercase letters represent matrices. For a complex matrix \(\mathbf{A}\in \mathbb{C}^{M\times N}\), its transpose, Hermitian, and inverse are denoted by \(\mathbf{A}^T,\mathbf{A}^H\) and $\mathbf{A}^{-1}$, respectively. Furthermore, \(x!\), \(\int x\,dx\) represent the factorial and integral operations of a real variable \(x\), respectively.  \(\sum_{i=1}^{N} a_i\) and \(\prod_{i=1}^{N} b_i\) denote the sum and product of a series of variables, respectively. For a random variable \(X\), \(\mathbb{E}[X]\) and \(\mathbb{P}(X>T)\) denote its expectation and the probability of the event \(X>T\), respectively. \(X\sim \mathcal{CN}(\mu,\sigma^2)\), \(X\sim \Gamma(\alpha,\beta)\) and \(X\sim \exp(\lambda)\) indicate the circularly symmetric complex Gaussian (CSCG), Gamma and exponential distributions for the random variable \(X\), respectively, where, \(\mu\) and \(\sigma^2\) represent the mean and variance for the CSCG distribution, \(\alpha\) and \(\beta\) are the shape and scale parameters for the Gamma distribution, and \(\lambda\) is the rate parameter for the exponential distribution. The Beta function is defined as \(B(a,b)=\int_{0}^{1}t^{a-1}(1-t)^{b-1}dt\), and the incomplete Gamma function is defined as \(B(x,a,b)=\int_{0}^{x}t^{a-1}(1-t)^{b-1}dt\). For a positive integer, the Gamma function satisfies \(\Gamma(x)=(x-1)!\).

\section{SYSTEM MODEL}
\subsection{Cooperative ISAC Model}
In the considered system, each BS is equipped with \(M_t\) transmit antennas and \(M_r\) receive antennas, while each user is equipped with a single antenna. The locations of the BSs are modeled as a two-dimensional homogeneous Poisson point process (HPPP) \(\Phi_{b}\) with intensity \(\lambda\). Specifically, \(\Phi_{b} = \left\{ \mathbf{d}_{i} =[x_i,\, y_i]^T\in \mathbb{R}^{2},\, \forall i \in \mathbb{N}^{+} \right\}\), where \(\mathbf{d}_i\) represents the location of the $i$-th nearest BS to the origin, \(\left\| \mathbf{d}_i \right\|\) denotes the Euclidean distance from the origin. The communication user and sensing target are distinct entities, both assumed to be randomly and uniformly distributed within the spatial domain. Here $\mathbf{p}_t$ and $\mathbf{p}_u$ denote the positions of the communication user and the sensing target, respectively. Similar to \cite{6856159,6376184}, our communication performance analysis focuses on a typical user located at the origin. Similarly, the typical target is also assumed to be located at the origin, where we utilize the BSs' location distribution to analyze the distance between the sensing BS and the typical target. This approach, validated by Slivnyak’s theorem for HPPP frameworks \cite{chiu2013stochastic}, preserves generality while simplifying derivations.\par
\begin{figure}
    \centering
    \includegraphics[width=0.8\linewidth]{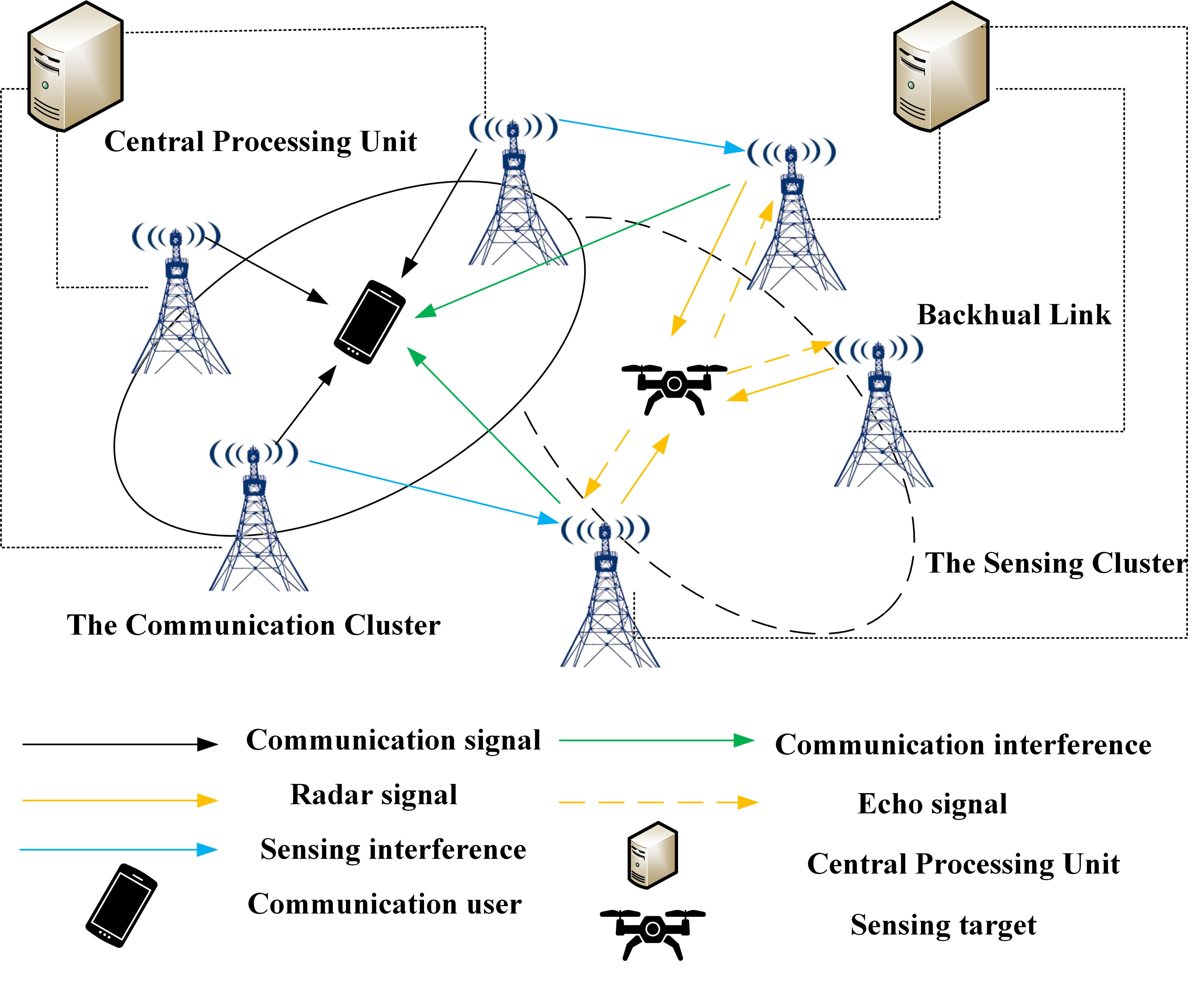}
    \caption{Cooperative sensing and communication for ISAC Network}
    \label{fig:system}
\end{figure}

To illustrate the cooperative communication and sensing framework in ISAC networks, we assume that each user is served by \(L\geq 1\) BSs, which form a communication cluster that employs CoMP transmission to deliver identical communication information signals. Similarly, each target is located by \(N\geq 1\) BSs, which constitute a sensing cluster, as shown in Fig. \ref{fig:system}. Each BS utilizes beamforming to simultaneously transmit a communication signal \(s^c\)  to its served user and a radar signal \(s^s\) toward the desired target. We assume that there is no correlation between the communication signal and the radar signal. The $i$-th BS transmits the signal \(\mathbf x_i=\mathbf{W}_i\mathbf{s}_i\in \mathbb{C}^{M_t \times 1}\), where \(\mathbf{W}_i=[\sqrt{p^s}\mathbf{w}_i^s,\sqrt{p^c}\mathbf{w}_i^c]\in \mathbb{C}^{M_t \times 2}\) denotes the beamforming matrix, with \(p^s\) and \(p^c\) representing the transmit power allocated to sensing and communication, respectively. The desired transmitted signal vector is defined as \(\mathbf{s}_i=[s_i^s,s_i^c]^T\in \mathbb{C}^{2\times 1}\), and we have \(\mathbb{E}[\mathbf{s}_i\mathbf{s}_i^H]=\mathbf{I}_2\). We assume that channel state information (CSI) is available, which can be obtained through efficient channel estimation algorithms. To mitigate the interference between sensing and communication and to reduce precoding complexity, zero-forcing (ZF) beamforming is adopted. The specific form of the $i$-th BS beamforming matrix is given as follows:
\begin{equation}
    \mathbf{W}_i=\mathbf{H}_i^H (\mathbf{H}_i\mathbf{H}_i^H)^{-1},\label{eq:1}
\end{equation}
where \(\mathbf{H}_i=[\left(\mathbf{a}^H(\theta_i)\right)^T,\left(\mathbf{h}^H_{i,c}\right)^T]^T\in \mathbb{C}^{2\times M_t }\). \(\mathbf{a}(\theta_i) \in \mathbb{C}^{M_t\times 1}\) denotes the sensing channel between the $i$-th BS and the typical target, and \(\mathbf{h}_{i,c}\in \mathbb{C}^{M_t\times 1}\) denotes the communication channel between the $i$-th BS and the typical user.
\subsection{Cooperative Communication Model}
Assuming that each user is served by \(L\) nearest BSs, and the corresponding set of serving BSs is denoted as $\Phi_a$. The received signal at the typical user is expressed as: 
\begin{equation}
\begin{aligned}
y_c&=\sum_{i=1}^{L}\|\mathbf{d}_i\|^{-\frac{\beta}{2}}\mathbf{h}_i^H\mathbf{W}_i\mathbf{s}_i+\sum_{j\in\Phi_b\setminus\Phi_a}\|\mathbf{d}_j\|^{-\frac{\beta}{2}}\mathbf{h}_j^H\mathbf{W}_j\mathbf{s}_j+n_c,
\end{aligned}
\end{equation}
where \(\beta\) is the path loss exponent ($\beta>2$). Without loss of generality, \(n_c\) denotes the additive white Gaussian noise (AWGN) with zero mean and unit variance. \(\|\mathbf{d}_i\|\) denotes the Euclidean distance between the typical user and the $i$-th nearest BS within the cooperative communication cluster, and \(\mathbf{h}_i\sim\mathcal{CN}(0,\mathbf{I}_{M_t})\) is the corresponding channel vector. \(\|\mathbf{d}_j\|\) denotes the distance between the typical user and the $j$-th interfering BS located outside the cooperative communication cluster.\par
Owing to the dominance of interference over noise in this ISAC scenario, our analysis focuses on the interference-limited condition, with the noise impact being neglected. The communication signal-to-interference ratio (SIR) is defined as: 
\begin{equation}
    SIR_c=\frac{\sum_{i=1}^Lp^c g_i\|\mathbf{d}_i\|^{-\beta}}{\sum_{j\in\Phi_b\setminus\Phi_a}p^tg_j\|\mathbf{d}_i\|^{-\beta}},
\end{equation}
where \(p^t=p^c+p^s\) represents the total transmit power, normalized to \(p^t=1\) for analytical simplicity. Here, \(g_i=|\mathbf{h}_i^H\mathbf{w}_i^c|^2\) denotes the small-scale fading parameter for desired signal power, while \(g_j=p^c|\mathbf{h}_j^H\mathbf{w}_j^c|^2+p^s|\mathbf{h}_j^H\mathbf{w}_j^s|^2\) characterizes the small-scale fading for the interference power. This work prioritizes the analysis of the communication coverage probability, which serves as a critical metric for evaluating the system’s interference resistance capability. In general, a higher coverage probability under a given threshold indicates stronger resistance to interference. The coverage probability is mathematically defined as follows:
\begin{equation}
    \mathcal{P}_c=\mathbb{P}(SIR_c\geq T),
\end{equation}
where \(T\) denotes the SIR threshold for communication coverage. Cooperative communication is considered to be successful when the received SIR at the user exceeds the predefined threshold, consistent with the definition provided in \cite{goldsmith2005wireless}. 
\subsection{Cooperative Sensing Model}
Assuming that each target is accurately located by $N$ nearest BSs, and the corresponding cooperative BSs set is denoted as $\Phi_s$. We focus on the echo signal reflected by the typical target and received by the closest sensing BS, which is labeled as BS 1, to simplify the analysis. We neglect the interference from the line of sight between the echo signal receiving BS and cooperative sensing BSs, because this interference can be eliminated by some techniques \cite{10736660}. The echo signal is expressed as follows:
\begin{equation}
    \begin{aligned}
y_1& = \sum_{i=1}^N \mathbf{v}(\theta_1) \sigma \|\mathbf{d}_1\|^{-\frac{\beta}{2}} \mathbf{b}(\theta_1) \|\mathbf{d}_i\|^{-\frac{\beta}{2}} \mathbf{a}^H(\theta_i)p^s \mathbf{w}_i^s s_i^s \\
&+ \sum_{q \in \Phi_b \setminus \Phi_s} \mathbf{v}(\theta_1) \|\mathbf{d}_1-\mathbf{d}_q\|^{-\frac{\beta}{2}}\mathbf{H}_{1,q}\mathbf{x}_q,
    \end{aligned}
\end{equation}
where \(\mathbf{v}(\theta_1)=\frac{1}{\sqrt{M_r}}[1,\,\cdots,\,e^{j\pi (M_r-1)\cos(\theta_1)}] \in\mathbb{C}^{1\times M_r}\) denotes the combining vector at the receiving side. \(\theta_1\) is the angle of arrival (AOA) of the echo signal and \(\sigma\) denotes the radar cross section (RCS) of the target, which can be estimated from the prior information. The steering vectors are defined as follows: the transmit steering vector is \(\mathbf{a}^H(\theta_1)=[1,\cdots,e^{-j\pi (M_t-1)\cos(\theta_1)}]\in \mathbb{C}^{1\times M_t}\), and the receive steering vector is \(\mathbf{b}(\theta_1)=[1,\cdots,e^{-j\pi (M_r-1)\cos(\theta_1)}]^T \in \mathbb{C}^{M_r\times 1}\). \(\|\mathbf{d}_1-\mathbf{d}_q\|\) denotes the distance between interfering BS $q$ and BS $1$, and \(\mathbf{H}_{1,q}\in \mathbb{C}^{M_r \times M_t}\) is the corresponding channel matrix, where each element independently follows a complex Gaussian distribution with zero mean and unit variance.\par
Analogous to the use of the user information rate as a performance metric in communication systems, we adopt the radar information rate as a measure of sensing performance for evaluating sensing effectiveness. Prior studies have demonstrated that the radar information rate is closely associated with radar capabilities, where a higher radar information rate corresponds to enhanced sensing performance. Since the radar information rate is logarithmically proportional to the sensing SIR, the sensing SIR needs to first be established before determining the radar information rate. The expression for the sensing SIR is given as follows:
\begin{equation}
    SIR_s=\sigma^2 M_r\frac{\|\mathbf{d}_1\|^{-\beta}\sum_{i=1}^Np^s f_i\|\mathbf{d}_i\|^{-\beta}}{\sum_{q\in \Phi_b \setminus \Phi_s}p^t f_q\|\mathbf{d}_1-\mathbf{d}_q\|^{-\beta}}
\end{equation} 
where \(f_i=|\mathbf{a}^H(\theta_i)\mathbf{w}_i^s|^2\) denotes the small-scale fading parameter associated with the effective sensing echo signal power, and \(f_q=p^s|\mathbf{v}(\theta_1)\mathbf{H}_{1,q}\mathbf{w}_q^s|^2+p^c|\mathbf{v}(\theta_1)\mathbf{H}_{1,q}\mathbf{w}_q^c|^2\) represents the small-scale fading parameter corresponding to the interference signal power. The radar information rate is given by
\begin{equation}
    R_s=\mathbb{E}[\log(1+SIR_s)].
\end{equation}
\section{PERFORMANCE ANALYSIS}
This section derives analytical expressions for cooperative ISAC performance metrics, namely communication coverage probability and radar information rate, by using stochastic geometry and probability theory. Furthermore, we provide closed-form expressions for some special cases, yielding valuable insights on the BS deployment strategies.
\subsection{Communication Performance Analysis}\label{communication}
In this subsection, we derive an analytical expression for the communication coverage probability in multi-BS cooperative communication networks. This derivation is challenging due to the involvement of multiple random variables, which introduce both analytical complexity and inherent stochasticity into the expression, thereby necessitating the implementation of simplified assumptions to address the challenge. Following the methodology in \cite{10769538}, we initially assume the distances between the served user and the BSs within the cooperative communication cluster as known. These distances are formally defined as \(r_1=\|\mathbf{d}_1\|\), \(r_2=\|\mathbf{d}_2\|\), \(\cdots\) ,\(r_L=\|\mathbf{d}_L\|\). Building upon these distances, we derive the conditional communication coverage probability. Nevertheless, the structural complexity in the numerator of the SIR expression poses considerable analytical challenges to this derivation. In Conjecture \ref{lemma1}, we propose an approximation method for the small-scale fading parameters to simplify the numerator.\par
\begin{figure}
    \centering
    \includegraphics[width=0.8\linewidth]{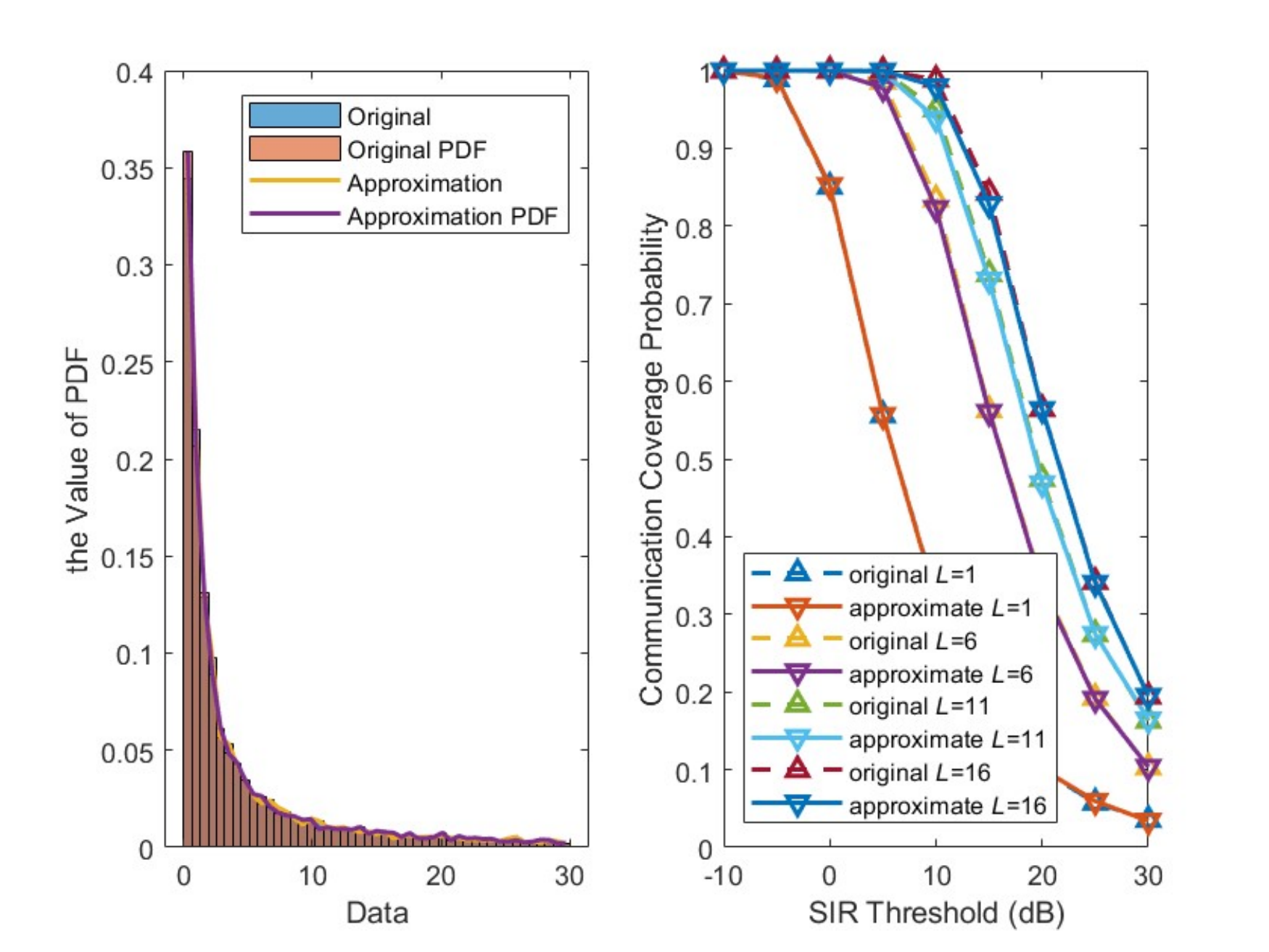}
    \caption{Verifications for the proposed approximation in Conjecture \ref{lemma1}. (a) The comparison of the PDF between original and approximate results. (b) The comparison of the communication coverage probability between original and approximate results}
    \label{fig:g_approximation}
\end{figure}

\begin{conjecture}\label{lemma1}
     Let random variable X \(\sim\) \(\Gamma(N,1)\), and let R denote random distance between the origin and a point located on the same plane. The finite sum of the products of two random variables, \(\sum_{i=1}^{L}x_ir_i^{-\alpha}\), can be approximately regarded as \(x\sum_{i=1}^Lr_i^{-\alpha}\), \(\alpha\) is a positive constant.
\end{conjecture}
The proof of Conjecture \ref{lemma1} is conducted by comparative analysis of the probability density functions (PDFs) of two sets of random variables. Furthermore, we assess the robustness of the proposed approximation across varying numbers of cooperative communication BSs $L$. As depicted in Fig. \ref{fig:g_approximation}, the close alignment between both the PDFs and the coverage probability of the original results with the approximate values demonstrates the precision and validity of our method. The validity of Conjecture \ref{lemma1} arises from the dominance of the distance term, which is governed by the exponential parameter $\alpha$, over the Gamma-distributed random variable. Conjecture \ref{lemma1} provides an approximation of the small-scale fading parameters, thereby improving the analytical tractability of the derivation. \par
By applying the approximation in Conjecture \ref{lemma1}, we obtain an expression of the cumulative distribution function (CDF) of a Gamma-distributed random variable. Previous studies have tackled similar challenges by employing bounding approximation techniques \cite{10556618,6932503}. However, experimental observations indicate that the approximation method shows a limited alignment with the theoretical CDF, while superior agreement can be attained via alternative approaches. To this end, we propose an enhanced methodology, provided in Conjecture \ref{approximation} to improve the accuracy of the theoretical CDF. 
\begin{conjecture}\label{approximation}
   For a normalized Gamma random variable \(g\), we derive an approximate CDF \(G(\gamma)\) using the Kolmogorov-Smirnov (K-S) statistic \(D\), and express it as follows:
\begin{equation}
    G(\gamma)=\mathbb{P}(g<\gamma)\approx \left[1-e^{-\alpha \gamma}\right]^N,
\end{equation}
where \(g\sim \Gamma(N,\frac{1}{N})\), \(\gamma\) is a positive constant and \(\alpha\) is a tunable parameter used to control the approximation accuracy between the analytical and real CDFs. The optimal \(\alpha\)  is determined through an iterative optimization process that minimizes discrepancy metric \(D\), formulated as:
\begin{equation}
    \alpha=\min \,D,
\end{equation}
where \(D=\max\,|F(\gamma)-G(\gamma)|\) is the K-S statistic, \(F(\gamma)=\left[1-e^{-\alpha \gamma}\right]^N\) is the analytical approximation of the CDF $G(\gamma)$.
\end{conjecture}
\begin{figure}
     \centering
     \includegraphics[width=0.8\linewidth]{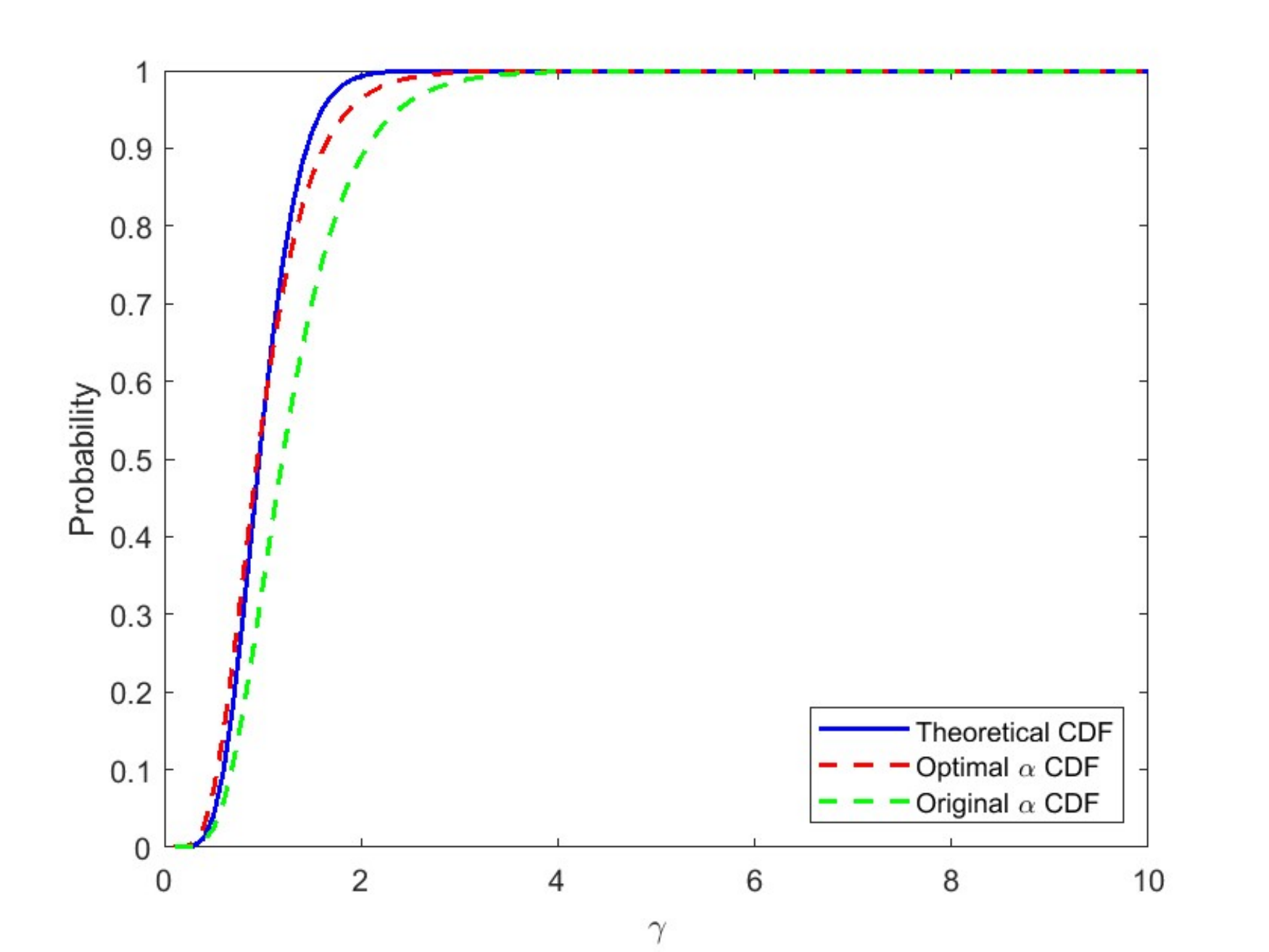}
     \caption{Verification for the optimal \(\alpha\) approximation}
     \label{optimal_alpha}
 \end{figure}
The proof of Conjecture \ref{approximation} is established by evaluating the discrepancy between empirical and theoretical CDF curves. As shown in Fig. \ref{optimal_alpha}, a comparison between the CDF curve generated and the optimal $\alpha$ with the original curve from prior works \cite{10556618,6932503} demonstrates that the optimized version achieves a significantly closer alignment with the theoretical CDF. \par
Building on the results established in Conjecture \ref{lemma1} and Conjecture \ref{approximation}, Theorem \ref{Theorem 1} rigorously derives the final analytical expression for the communication coverage probability. 
\begin{theorem}\label{Theorem 1}
    The communication coverage probability can be derived as:
\begin{equation}
\begin{aligned}
    \mathcal{P}_c &= \mathbb{P}(SIR_c \geq T) \\
    &= \sum_{n=1}^{Q}(-1)^{n+1}\binom{Q}{n}\int_{0}^{\infty}\cdots \int_{0}^{\infty}e^{-\pi\lambda H_1(Q,n,\beta,T,p_t,p_c)}\\
       &\quad \times f(r_1)\cdots f(r_L)dr_1\cdots dr_L,
\end{aligned}
\end{equation}
where \(T\) is the communication SIR threshold, \(Q\triangleq M_t-1\), \(Q_L\triangleq\left(1+\frac{\alpha nTp^tr_L^{-\beta}}{DQp^c}\right)^{-1}\), \(H_1(Q,n,\beta,T,p_t,p_c)\triangleq\)
\begin{equation*}
    \frac{2}{\beta}\left(\frac{\alpha nTp^t}{QDp^c}\right)^{\frac{2}{\beta}}\left(B(\frac{2}{\beta},1-\frac{2}{\beta})-B(Q_L,\frac{2}{\beta},1-\frac{2}{\beta})\right).
\end{equation*}
\end{theorem}
\begin{IEEEproof}
Please refer to Appendix~A.
\end{IEEEproof}
Although the expression in Theorem \ref{Theorem 1} provides a general analytical form for the communication coverage probability, its mathematical complexity obscures the underlying physical interpretation. A more tractable expression can be derived for specific parameter configurations. Guided by this finding, we derive a closed-form solution for the communication coverage probability in Proposition \ref{popostion_1}.
\begin{proposition}\label{popostion_1}
When $L=1$ , \(\beta=4\) , communication coverage probability \(\mathcal{P}_c \) can be given as follows
    \begin{equation}\label{closed}
        \mathcal{P}_c=\sum_{n=1}^{Q}(-1)^{n+1}\binom{Q}{n}\frac{1}{1+\left(\frac{T\alpha np^t}{Qp^c}\right)^{\frac{1}{2}}\left(\frac{\pi}{2}-\arcsin(\sqrt{U_L'})\right)},
    \end{equation}
    where \(U_L'\triangleq\left(1+\frac{T\alpha np^t}{Qp^c}\right)^{-1}\).
\end{proposition}
\begin{IEEEproof}
    Please refer to Appendix~B.
\end{IEEEproof}
Remarkably, the expression in \eqref{closed} is independent of the BS deployment density, indicating that the coverage probability exhibits invariance to variations of this parameter's value. 
\subsection{Sensing Performance Analysis}\label{sensing_performance}
\begin{figure}
    \centering
    \includegraphics[width=0.8\linewidth]{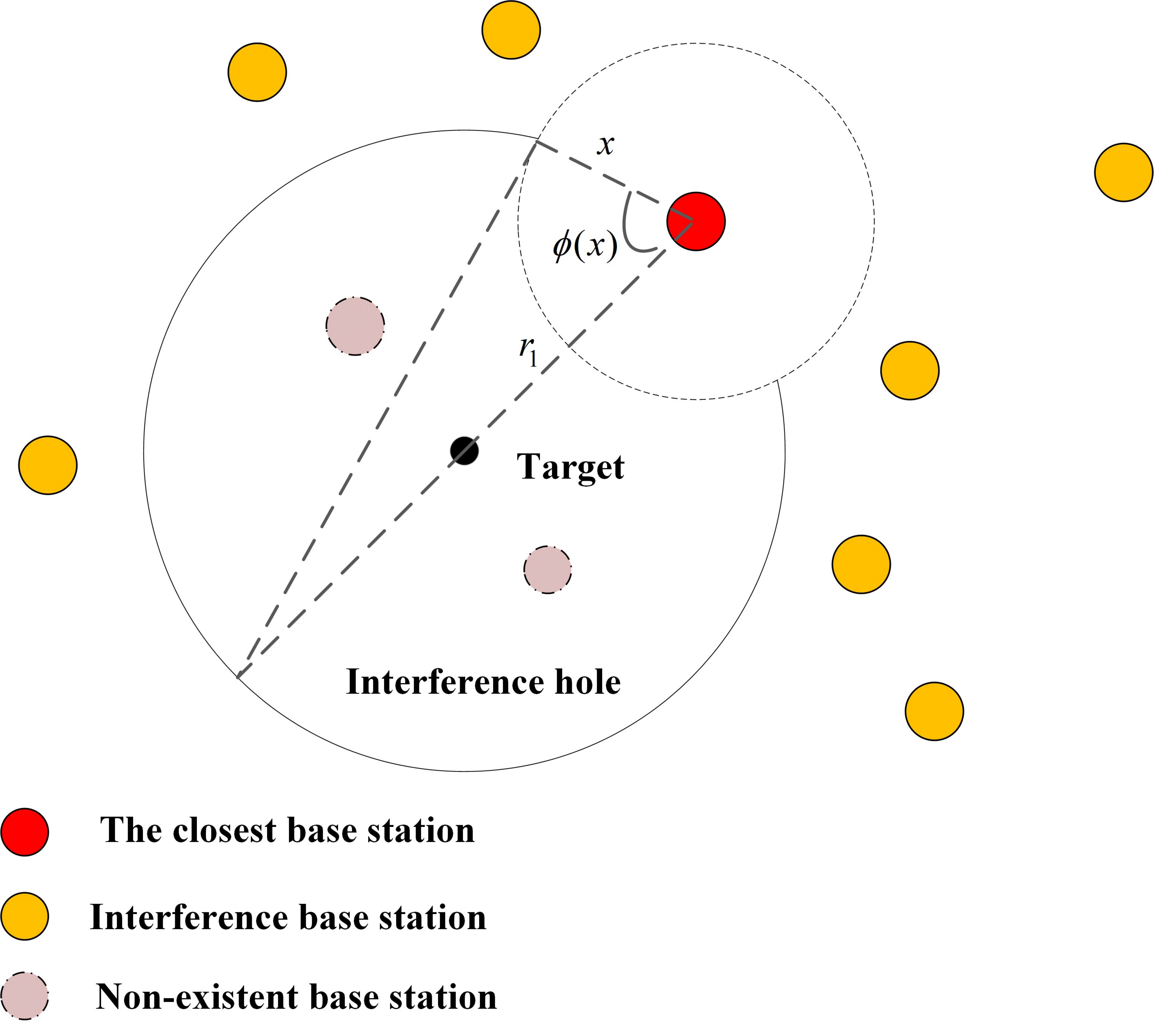}
    \captionsetup{justification=raggedright, singlelinecheck=false}
    \caption{Illustration of interference hole.}
    \label{fig:interference hole}
\end{figure}

In this subsection, we derive the expression for the radar information rate. Due to the independence of BS locations and small-scale fading parameters, the desired echo signal power is independent of the interference signal power. According to \cite{5407601}, the expectation of the logarithm of the ratio between two uncorrelated variables can be expressed in terms of an integral involving the Laplace transforms of these variables. Thus, the expected radar information rate is given by:
\begin{equation}
\begin{aligned}
    &\mathbb{E}[\log(1+SIR_s)]
    \\ =&\mathbb{E}\left[\log(1+\sigma^2 M_r\frac{\|\mathbf{d}_1\|^{-\beta}\sum_{i=1}^Np^s f_i\|\mathbf{d}_i\|^{-\beta}}{\sum_{q\in \Phi_b \setminus \Phi_s} p^tf_q\|\mathbf{d}_1-\mathbf{d}_q\|^{-\beta}})\right]
    \\=&\int_{0}^{+\infty} \frac{1-\mathbb{E} [e^{-zX}]}{z}\mathbb{E}[e^{-zY}]dz, \label{eq:rate}
\end{aligned}
\end{equation}
where \(X\,\triangleq\, \sigma^2 M_rp^s\sum_{i=1}^N f_i\|\mathbf{d}_i\|^{-\beta}\), and \(Y\, \triangleq\, \sum_{q\in \Phi_b \setminus \Phi_s }p^t f_q\|\mathbf{d}_1-\mathbf{d}_q\|^{-\beta}\|\mathbf{d}_1\|^{\beta}\). $\mathbb{E} [e^{-zX}]$ and $\mathbb{E} [e^{-zY}]$ are Laplace transform expressions for variable $X$ and $Y$, respectively. As shown in \eqref{eq:rate}, the derivation of the radar information rate translates to determining the Laplace transform expressions of desired echo signal power and those of interference signal power. We first derive expressions for the Laplace transform given the minimum and maximum distances within the cooperative sensing cluster \(r_1=\|\mathbf{d}_1\|\), \(r_N=\|\mathbf{d}_N\|\), as well as the distance ratio \(\eta_N=\frac{r_1}{r_N}\). Rigorous analytical results for the conditional Laplace transform expressions are presented in Lemma \ref{lemma_sen}.
\begin{lemma}\label{lemma_sen}
Given the minimum BS distance \(r_1\), maximum BS distance within the cooperative sensing cluster \(r_N\), and the distance ratio \(\eta_N\), the Laplace transforms corresponding to the desired echo signal power and interference signal power are formulated as follows:\par
\begin{align}
    & \mathbb{E} \left[e^{-zX}\right|r_N\,] = \exp\left(-\frac{2\pi\lambda}{\beta}  H_3(M_t,z,\sigma^2,M_r,r_N,\beta,p^s)\right), \\
    & \mathbb{E}\left[e^{-zY}|r_1,\eta_N\right] = \exp\left(-2\pi\lambda r_1^2 H_4(z,\beta,\eta_N)\right),
\end{align}
where $H_3(M_t,z,\sigma^2,M_r,r_N,\beta,p^s)$ and $H_4(z,\beta,\eta_N)$ can be written as
\begin{equation}
    \begin{aligned}
        & H_3(M_t,z,\sigma^2,M_r,r_N,\beta,p^s) \triangleq\\
        & (z\sigma^2M_rp^s)^{\frac{2}{\beta}} \sum_{i=1}^{Q}\binom{Q}{i}B\left(u_N,Q-i+\frac{2}{\beta},i-\frac{2}{\beta}\right), \\
        & H_4(z,\beta,\eta_N)\triangleq \\
        & \frac{1}{\beta}z^{\frac{2}{\beta}}\left(B\left(\frac{2}{\beta},1-\frac{2}{\beta}\right)-B\left(v_N,\frac{2}{\beta},1-\frac{2}{\beta}\right)\right),
    \end{aligned}
\end{equation}
with the following definitions:
\(Q\triangleq M_t-1\), $u_N\triangleq(p^s z \sigma^2 M_r r_N^{-\beta} + 1)^{-1}$, \(v_N\triangleq\frac{1}{1+z\eta_N^{\beta}}\), and the distance ratio $\eta_N=\frac{r_1}{r_N}$.
\end{lemma}
\begin{IEEEproof}
    Please refer to Appendix~C.
\end{IEEEproof}
Building on the conditional Laplace transform expressions in Lemma \ref{lemma_sen}, we derive the analytical expression for the expected radar information rate in Theorem \ref{theore_sen}.
\begin{theorem}
    The radar information rate is derived as follows:
\begin{equation}
\begin{aligned}
        &R_s \\
    &= \int_{0}^{+\infty} \frac{1}{z}\left(1 - \int_0^{\infty} e^{-\frac{2\pi\lambda}{\beta} H_3(M_t,r_N,\beta,p^s)}  f(r_N)  \, dr_N\right)  & \\ 
&\quad \times \int_0^{1} \frac{f(\eta_N)}{1 + 2 H_4(z,\beta,\eta_N)}  \, d\eta_N dz,
\end{aligned}    
\end{equation}  \label{theore_sen}
where \(f(r_N)=\frac{2(\lambda \pi r_N^2)^{N}}{\Gamma(N)r_N}e^{-\lambda \pi r_N^2}\), \(f(\eta_N)=2(N-1)\eta_N(1-\eta_N^2)^{N-2}\).
\end{theorem}
\begin{IEEEproof}
    Please refer to Appendix~D.
\end{IEEEproof}
Interference hole refers to a phenomenon in which, given the closest distance between the target and the BS, the distribution of interfering BSs is not completely random, but there is a restricted region devoid of interfering BSs. This region is depicted as a circular region with a radius of $r_1$, centered at the target, as illustrated in Fig. \ref{fig:interference hole}. In collaborative sensing mode, the impact of the interference hole is not considered, due to the low probability that interfering BSs are located within the interference hole. However, under the non-collaborative sensing mode, the impact must be accounted for, as the likelihood of interfering BSs being located within the interference hole increases significantly, thereby resulting in deviations of up to 15\% from the actual value \cite{10735119}. We present the detailed analysis in Proposition \ref{Proposition_sens}.
\begin{proposition} \label{Proposition_sens}
    When $N=1$, the radar information rate \(R_s\) is given as follows
\begin{equation}\label{hole_expression}
    \begin{aligned}
           R_s=&\int_{0}^{\infty}\frac{1}{z}\left(1-\frac{1}{(1+z\sigma^2M_rp^s)^{M_t-1}}\right)
        \\ &\phantom{=}\int_{0}^{\infty} f(r_1)\exp\left(-\pi\lambda r_1^{4} \frac{2}{\beta} (zp^t)^{\frac{2}{\beta}} 
            B\left(\frac{2}{\beta}, 1-\frac{2}{\beta}\right)\right) \\
             &\phantom{=}\times \exp\left(\lambda r_1^2 \int_{0}^{2} 2\arccos{\frac{t}{2}} 
            \frac{zp^t  r_1^{\beta} t^{-\beta}}{1+zp^t  r_1^{\beta} t^{-\beta}} t\, dt\right)\,dz,
    \end{aligned}
\end{equation}
where \(f(r_1)=2\pi\lambda e^{-\pi\lambda r_1^2}\) is the PDF of the distance between the target and BS 1. 
\end{proposition}
\begin{IEEEproof}
    Please refer to Appendix~E.
\end{IEEEproof}
The expression in \eqref{hole_expression} takes into account the real distribution of the interference BSs, addressing the challenge of underestimation of the sensing SIR, thus improving the accuracy of the theoretical expression of the radar information rate.
\section{SIMULATIONS}
We employ Monte Carlo simulations to validate the accuracy of the derived expressions and to further explore how system performance is impacted by BS deployment density and other parameters. The deployment of random positions for the BSs and the realizations of the small-scale fading parameters are generated in each round of Monte Carlo simulations. The communication coverage probability and radar information rate are calculated through over $10^{6}$ simulation trials. The simulation parameters are set as follows: the number of transmit antennas is set to \(M_t=10\), and the number of receive antennas is \(M_r=6\), forming a multiple-antenna system that provides enhanced spatial diversity and multiplexing gains. The path loss factor is set to \(\beta=4\), the total transmit power is $p^t=1\,W$ at each BS. Furthermore, the average RCS of the detected target is assumed to be \(\sigma^2=1\).
\subsection{Communication Performance}
\begin{figure}
        \centering
        \includegraphics[width=0.8\linewidth]{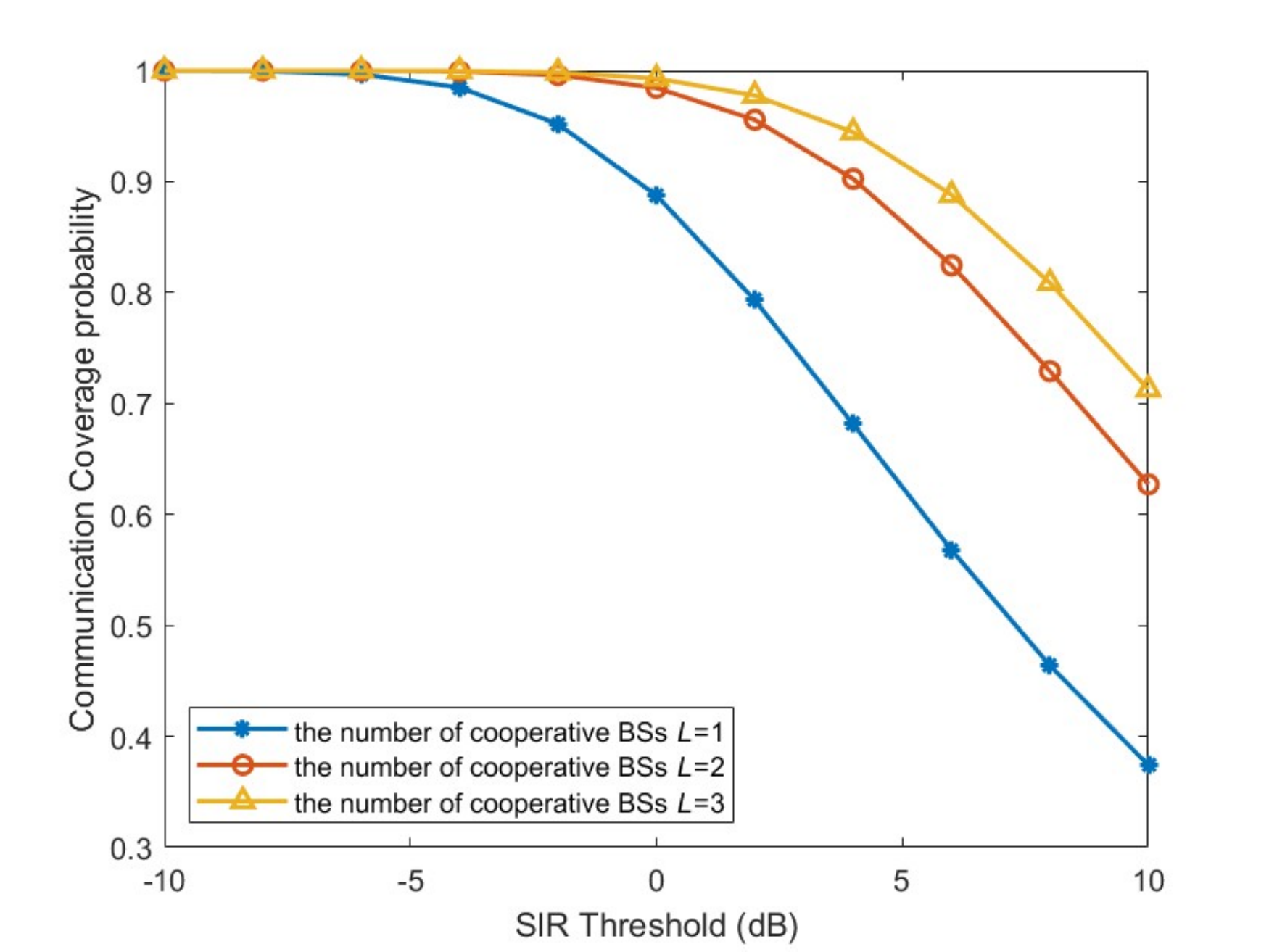}
        \caption{Communication coverage probability with respect to SIR Threshold with respect to $L$.}
        \label{fig:differnet_L}
    \end{figure}

Fig. \ref{fig:differnet_L} illustrates the relationship among the number of  cooperative BSs and SIR threshold and communication coverage probability. When the number of cooperative BSs is $L=2$, communication coverage probability shows a declining trend as the SIR threshold increases, indicating that achieving higher coverage probability is dependent on favorable communication conditions. Notably, the coverage probability at any given SIR threshold increases with the number of cooperative BSs increasing, reflecting the stronger interference resilience. This is because the reliance on single BS signal power diminishes as the number of cooperative communication BSs increases. When the signal from individual BSs experiences severe fading or encounters intensified interference, other BSs within the cooperative cluster can maintain stable communication links. Consequently, the received signal strength at the user terminal remains relatively consistent, thereby improving interference resistance capabilities. This finding highlights the practical advantages of cooperative ISAC systems in improving communication reliability under interference-limited conditions.\par
\begin{figure}
    \centering
    \includegraphics[width=0.8\linewidth]{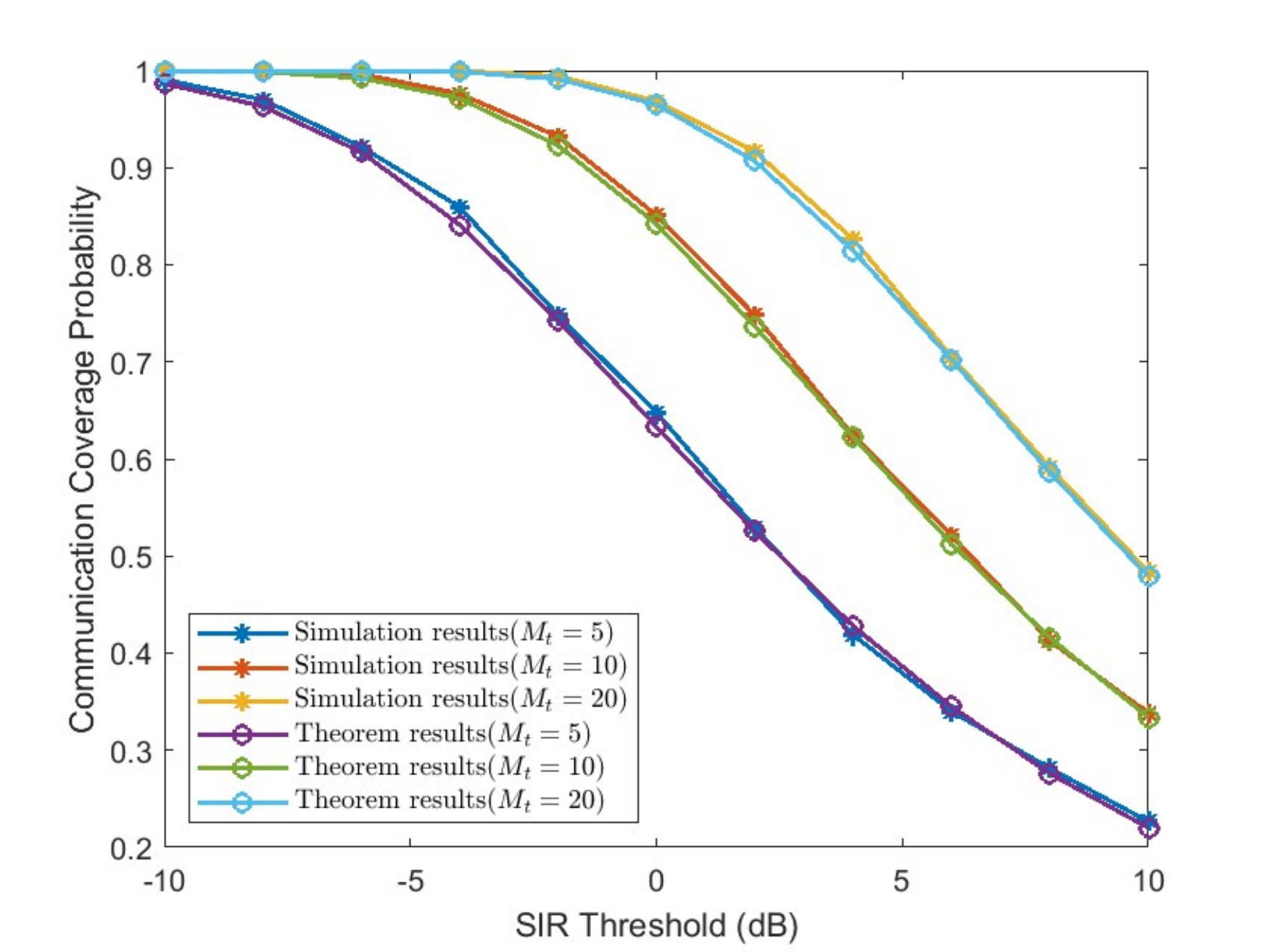}
    \caption{Communication coverage probability of optimal \(\alpha\) with respect to SIR Threshold with $L$ = 1.}
    \label{L_1_alphaed}
\end{figure}

\begin{figure}
    \centering
    \includegraphics[width=0.8\linewidth]{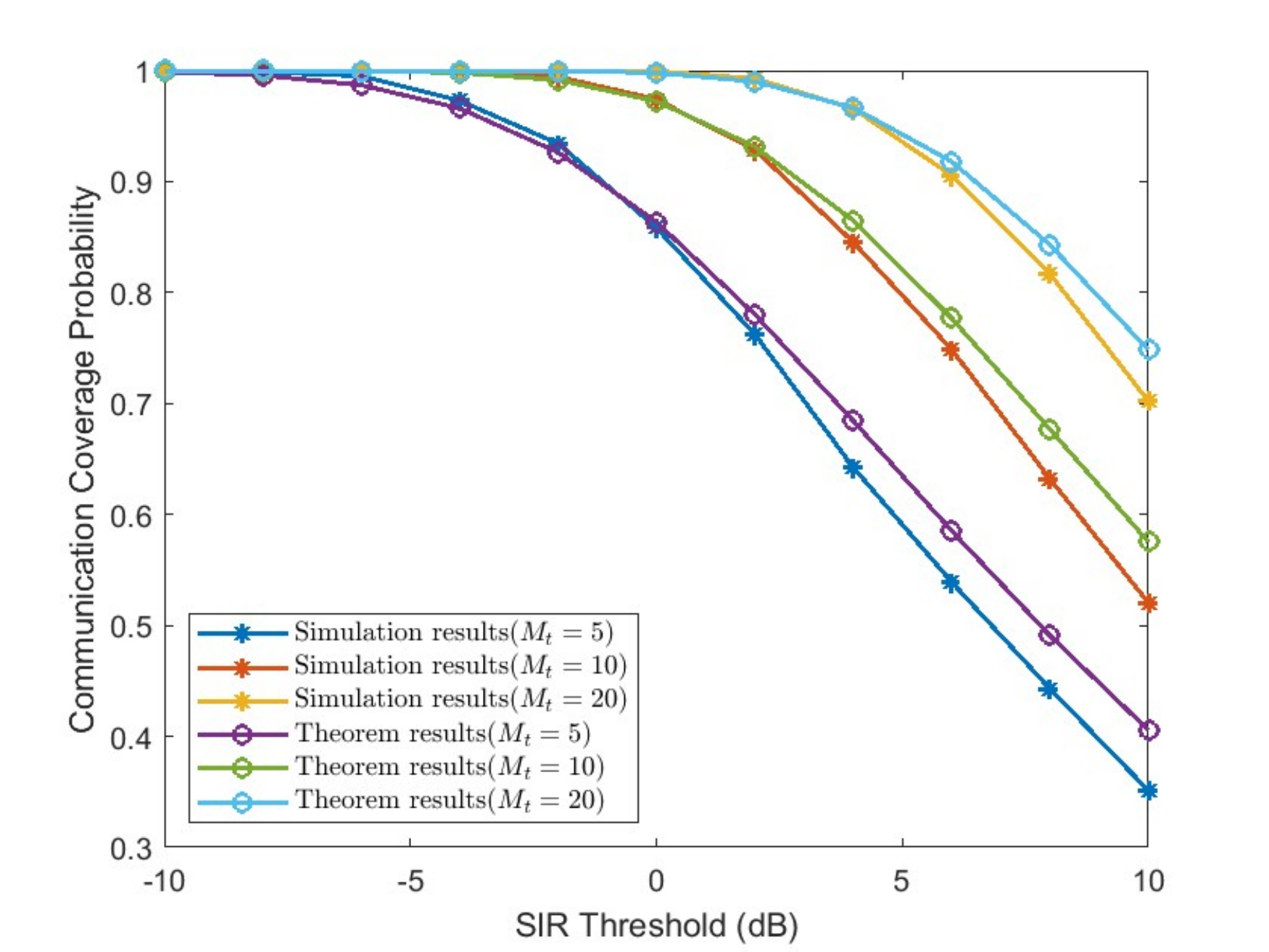}
    \caption{Communication coverage probability of optimal \(\alpha\) with respect to SIR Threshold with $L$ = 2.}
    \label{L_2_alphaed}
\end{figure}

Fig. \ref{L_1_alphaed} demonstrates that the results of the closed-form expression for $\mathcal{P}_c$ derived in Proposition \ref{popostion_1} aligns closely with the Monte Carlo simulation results, confirming the accuracy of our analysis. For any given SIR threshold, the coverage probability increases with the number of transmit antennas. This trend is predominantly driven by spatial diversity gains, which mitigate interference effects through multi-path signal exploitation, thereby improving the ability to resist interference. Fig. \ref{L_2_alphaed} shows the results of the analytical expression with the number of cooperative BSs set to 2 in Theorem \ref{Theorem 1} are consistent with simulation results. Similar to the trend observed in Fig. \ref{L_1_alphaed}, the incremental performance gains from increasing the number of transmit antennas diminish with this number increasing. Moreover, as the number of transmit antennas increases, system overhead also rises due to the increased complexity of signal processing and CSI estimation. Therefore, indiscriminately increasing the number of transmit antennas may not be an efficient strategy for enhancing system performance in cooperative ISAC networks.\par

\begin{figure}
    \centering
    \includegraphics[width=0.8\linewidth]{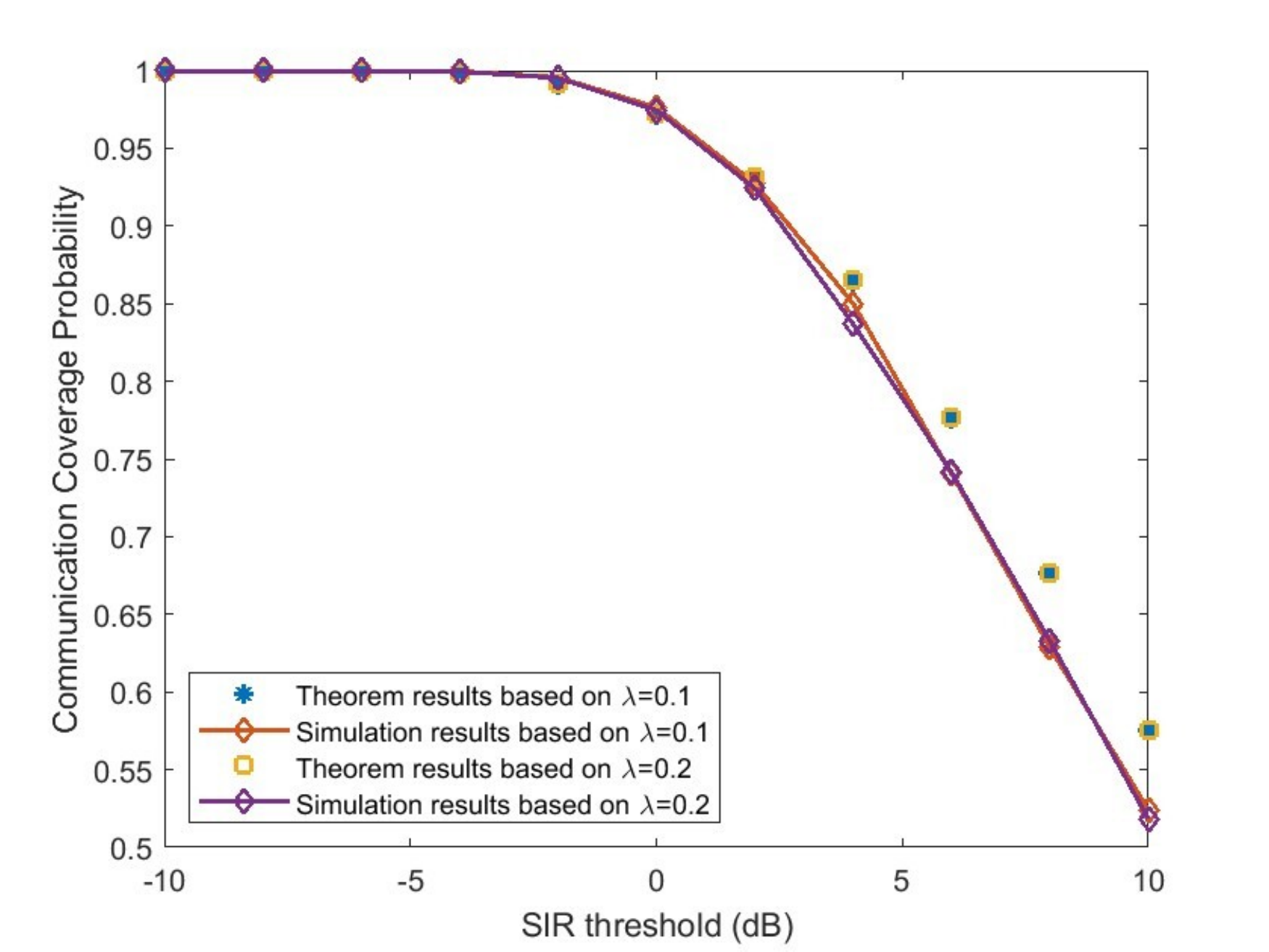}
    \caption{Communication coverage probability with respect to the BS deployment density.}
    \label{fig:the coverage probability with respect to the intensity of BS deployment}
\end{figure}

To investigate the impact of the BS deployment density on communication coverage probability, Fig. \ref{fig:the coverage probability with respect to the intensity of BS deployment} depicts the communication coverage probability versus the SIR threshold under scenarios with different BS deployment densities $\lambda$. It can be seen that both theoretical and simulation results exhibit the same phenomenon: changes in the BS deployment density have minimal impact on the coverage probability. Notably, the results are consistent with the analysis of Proposition \ref{popostion_1}. This phenomenon primarily stems from the tradeoff introduced by the increasing number of BSs. On one hand, a higher BS deployment density increases the likelihood that cooperative communication BSs are positioned closer to the served user, thereby reducing path loss for the desired signal and improving the communication SIR. On the other hand, the total interference power grows with the number of BSs, which adversely impacts the SIR. These two opposing effects offset each other’s influence on the SIR, ultimately resulting in a negligible net impact on coverage probability. This finding highlights that simply increasing BS deployment density is not an effective strategy for improving communication performance.

\subsection{Sensing Performance} 
\begin{figure}
    \centering
    \includegraphics[width=0.8\linewidth]{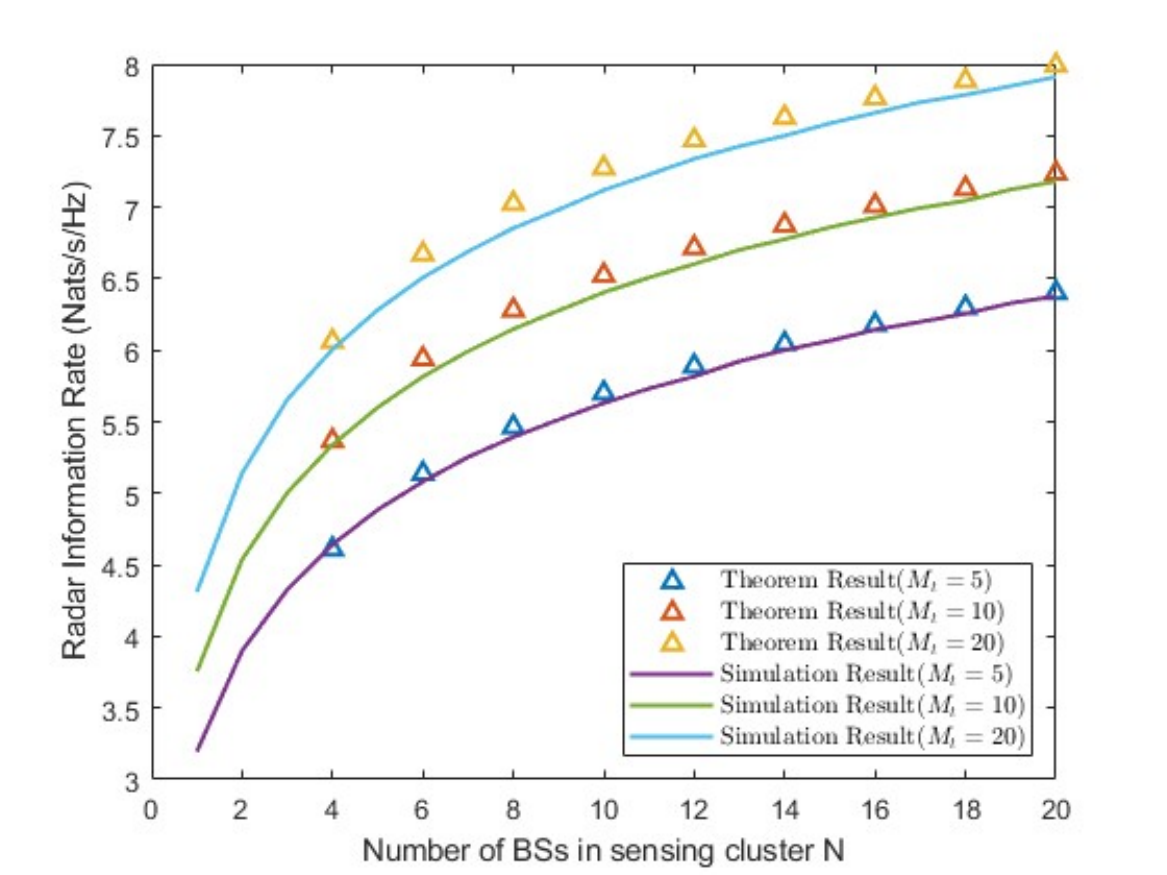}
    \caption{Radar information rate \(R_s\) with respect to number of BSs in sensing cluster $N$.}
    \label{fig:r_s vs N}
\end{figure}

Fig. \ref{fig:r_s vs N} presents the radar information rate \(R_s\) versus the number of BSs in cooperative sensing cluster under different numbers of transmit antennas. It is evident that the theoretical results derived in Theorem \ref{theore_sen} closely align with simulation results. As expected, \(R_s\) increases with the number of BSs due to multi-perspective sensing data fusion and enhanced acquisition of effective echo signals. Notably, while \(R_s\) 
improves significantly when the number of cooperative BSs is limited, the marginal performance gain diminishes as this number continues to rise. Furthermore, \(R_s\) exhibits significant improvement when the number of transmit antennas is relatively small, whereas the marginal gains diminish with further increases in this parameter. \par
\begin{figure}
    \centering
    \includegraphics[width=0.8\linewidth]{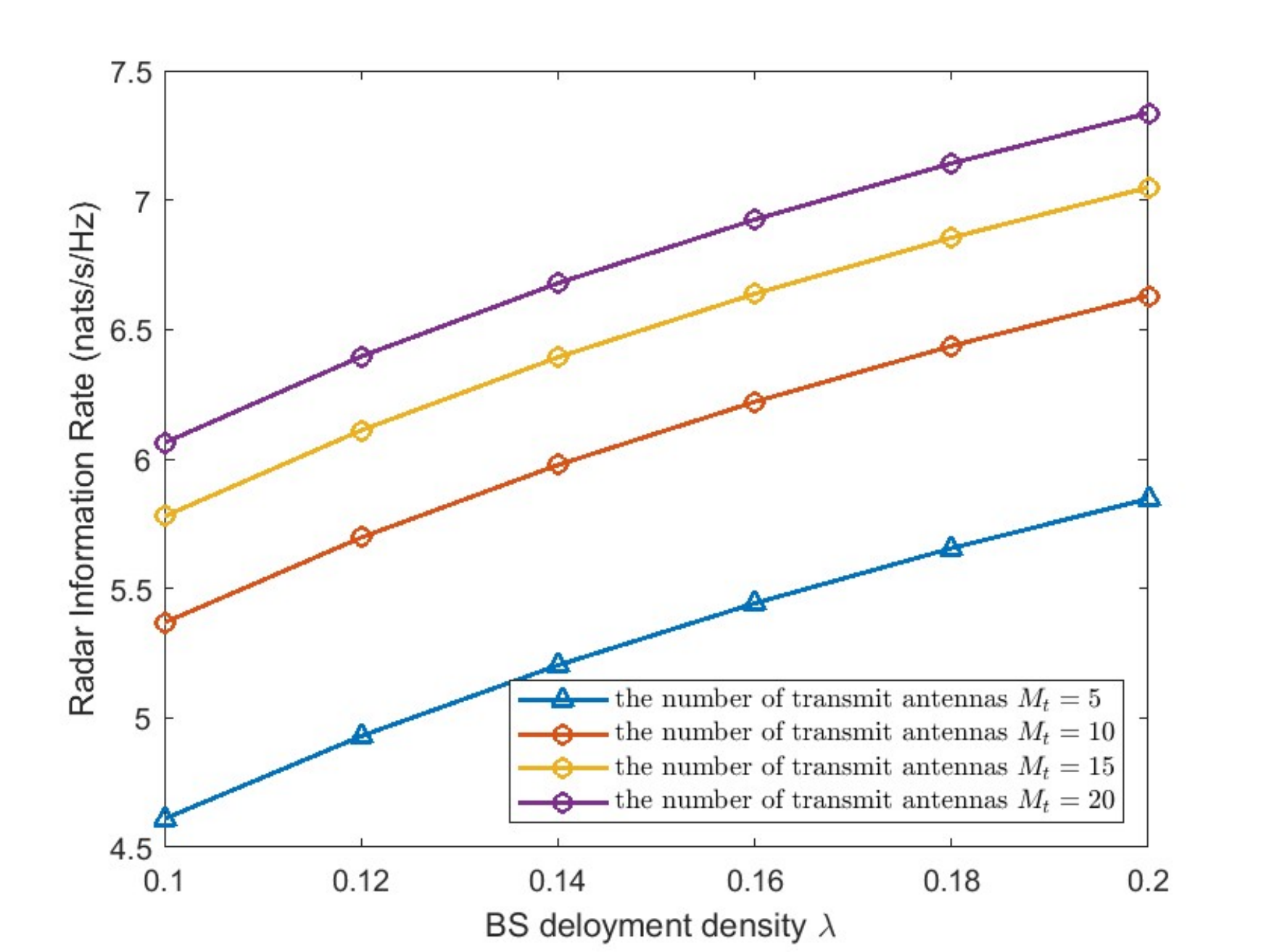}
    \caption{Radar information rate with respect to BS deployment density.} 
    \label{fig:the radar information rate with respect to the intensity of BS deployment}
\end{figure}
To explore the impact of the BS deployment density on the radar information rate, Fig. \ref{fig:the radar information rate with respect to the intensity of BS deployment} shows the radar information rate versus the BS deployment density $\lambda$. For any given number of transmit antennas, the radar information rate $R_c$ continues to increase with the BS deployment density $\lambda$. This improvement is primarily attributed to the increased likelihood that cooperative sensing BSs are located closer to the target as the BS density grows. A shorter propagation distance of the echo signal reduces power attenuation, resulting in enhanced echo signal power. Although more BSs also introduce additional interference, the enhanced echo signal power rather than accompanying interference is the predominant factor, ultimately enhancing the radar information rate.\par
\begin{figure}
    \centering
    \includegraphics[width=0.8\linewidth]{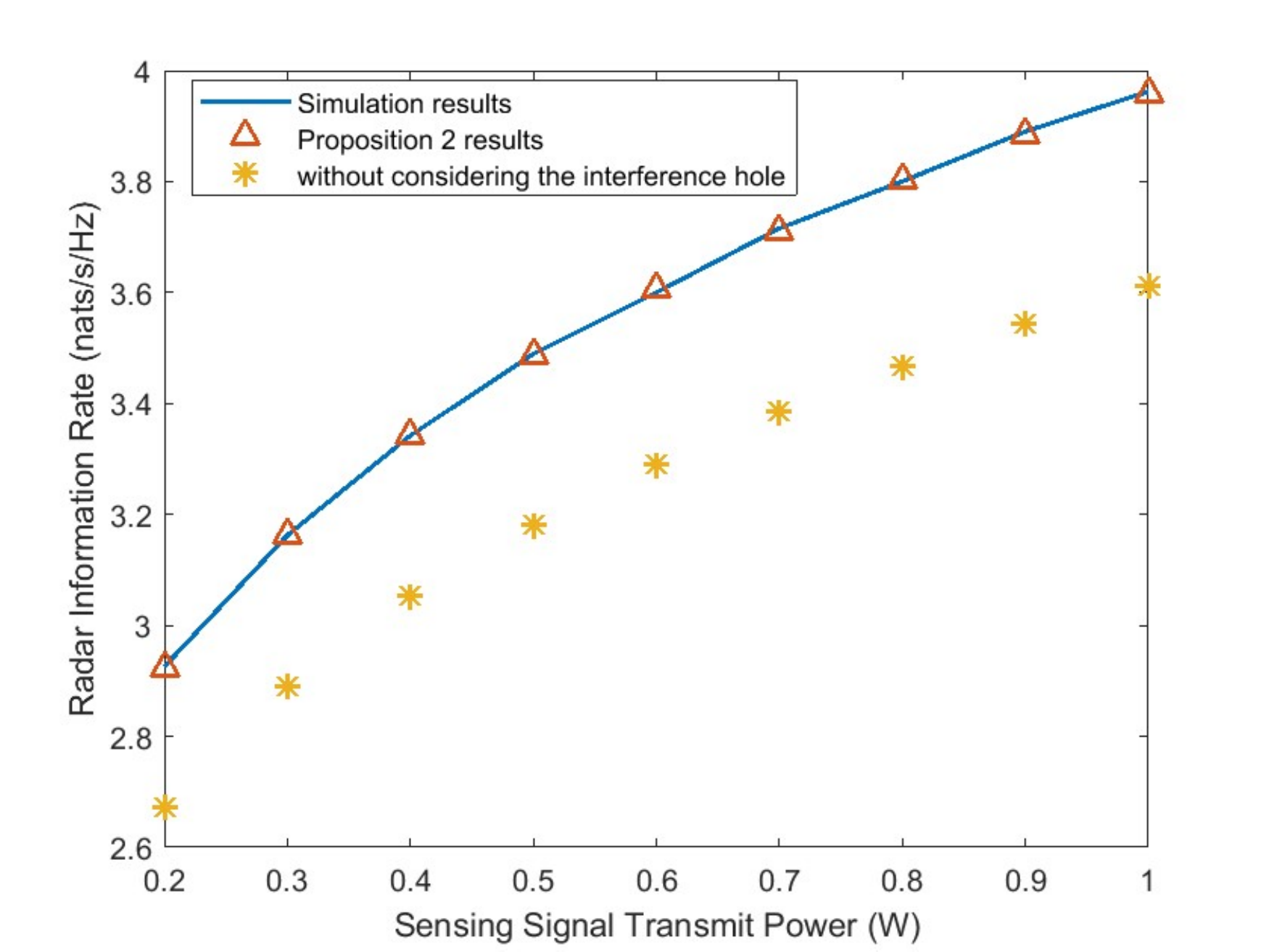}
    \caption{Radar information rate with respect to the sensing transmitting power.}
    \label{fig:sensing_interference_hole}
\end{figure}
Fig.\ref{fig:sensing_interference_hole} depicts the relationship between the radar information rate and the sensing signal transmit power. The results show that increasing transmit power improves radar sensing performance. The theoretical results derived in Proposition \ref{Proposition_sens} demonstrate close alignment with the Monte Carlo simulation results, which quantitatively validate the accuracy of the interference hole analysis. In addition, results obtained without considering the interference hole effect are included. The values are approximately $8\%-10\%$ lower than both the simulation and theoretical results. This discrepancy arises from the inclusion of interference contributions from non-existent interfering BSs. This overestimation of interference reduces the SIR at the echo-receiving BS, thereby causing a noticeable degradation in the radar information rate.\par
\section{CONCLUSION}
In this paper, we proposed a generalized stochastic geometry framework to model cooperative ISAC networks that incorporate CoMP transmission and multi-static radar sensing. We employed communication coverage probability and radar information rate as key performance metrics due to their critical role in evaluating interference resilience and radar detection capabilities. Based on stochastic geometry and probability theory, tractable analytical expressions for these metrics were derived for both general and specific conditions. The close agreement between the analytical expressions and Monte Carlo simulations was rigorously validated. Furthermore, we offered valuable insights into BS deployment strategies to optimize ISAC network performance. Increasing the number of cooperative BSs, rather than simply increasing the BS deployment density, proved to be a more effective strategy for improving communication coverage probability. Meanwhile, improvements in radar information rate can be achieved by increasing both BS deployment density and the number of cooperative BSs. Moreover, expanding the number of transmit antennas can improve both communication and sensing performance when the initial number of antennas is limited. Nevertheless, indiscriminate increases in the number of transmit antennas is not an efficient strategy for enhancing system performance in cooperative ISAC networks. 
\appendices
\section{Proof of Theorem 1}
According to the beamforming matrix definition of \eqref{eq:1}, the small-scale fading parameter for the desired signal power \(g_i\) follows a Gamma distribution \(\Gamma(M_t-1,1)\) \cite{6596082}. Meanwhile the small-scale fading parameter for the interference signal power \(g_j\) follows an exponential distribution \(\exp\,(1)\), which can be obtained using the moment matching method described in \cite{10769538}. As outlined in Section \ref{communication}, we first derive the conditional communication coverage probability:
\begin{equation}
    \begin{aligned}
        \mathcal{P}_c' &= \mathbb{P}(SIR_c \geq T|r_1,r_2,\cdots,r_L) \\
        &= \mathbb{P}\left(\frac{\sum_{i=1}^L p^cg_ir_i^{-\beta}}{\sum_{j\in\Phi_b\setminus\Phi_a} p^t g_j\|\mathbf{d}_j\|^{-\beta}} \geq T|r_1,r_2,\cdots,r_L\right)\\
        &\overset{(a)}{\approx}\mathbb{P}\left(\frac{g \quad p^c\sum_{i=1}^Lr_i^{-\beta}}{\sum_{j\in\Phi_b\setminus\Phi_a} p^t g_j\|\mathbf{d}_j\|^{-\beta}} \geq T|r_1,r_2,\cdots,r_L\right)\\
        &\overset{(b)}{=}\mathbb{P}\left(g \geq \frac{TI_c}{p^cD}|r_1,r_2,\cdots,r_L\right),\label{p_c}
    \end{aligned}
\end{equation}
where \(I_c=\sum_{j\in\Phi_b\setminus\Phi_a} p^tg_j\|\mathbf{d}_j\|^{-\beta}\) and \(D=\sum_{i=1}^L r_i^{-\beta}\) are interference signal strength and conditional distances, respectively. The approximation in step \((a)\)  is based on Conjecture \ref{lemma1}, and step \((b)\) is obtained by algebraic transformation of the inequality.\par
Given the small-scale fading parameter \(g \sim \Gamma\left(M_t-1,1\right)\), the normalized variable \(\frac{1}{M_t-1}g\) follows \(\Gamma\left(M_t-1,\frac{1}{M_t-1}\right)\). Consequently, the conditional communication coverage probability can be rewritten as follows:
\begin{equation}
    \begin{aligned}
        &\mathcal{P}_c'=\mathbb{P}\left(g \geq \frac{TI_c}{p^cD}|r_1,r_2,\cdots,r_L\right)\\
        &=1-\mathbb{P}\left(g \leq \frac{TI_c}{p^cD}|r_1,r_2,\cdots,r_L\right)\\
        &=1-\mathbb{P}\left(\frac{g}{Q} \leq\frac{TI_c}{Qp^cD}\right)\\
        &\overset{(a)}{\approx}1-\left[1-\exp{(-\alpha \gamma)}\right]^{Q}\\
        &\overset{(b)}{=}\sum_{n=1}^{Q}(-1)^{n+1}\binom{Q}{n}\mathbb{E}\left(e^{-\alpha n \frac{TI_c}{Qp^cD}}\right),\label{p_c_1}
    \end{aligned}
\end{equation}
where \(Q \triangleq M_t-1\), $\alpha$ is a constant obtained from Conjecture \ref{approximation} and \(\gamma\triangleq \frac{TI_c}{Qp^cD}\). The approximation in step \((a)\) is derived from the results based on Conjecture \ref{approximation}, and step \((b)\) follows from the application of the binomial theorem. Now we employ tools from stochastic geometry to calculate the  expectation term $\mathbb{E}\left(e^{-\alpha n \frac{TI_c}{Qp^cD}}\right)$ in \eqref{p_c_1} as
\begin{equation}\label{epec_com}
    \begin{aligned}
       & \mathbb{E}\left(e^{-\alpha n \frac{TI_c}{Qp^cD}}\right)\\
&= \mathbb{E}\left(e^{-\alpha n \frac{\sum_{j\in\Phi_b\setminus\Phi_a}T p^tg_j\|\mathbf{d}_j\|^{-\beta}}{Qp^cD}}\right)\\
&\overset{(a)}{=} \mathbb{E}\prod_{j\in\Phi_b\setminus\Phi_a}\mathbb{E}_{g_j} \left( e^{-\frac{\alpha n T p^tg_j\|\mathbf{d}_j\|^{-\beta}}{Qp^cD}}\right)\\
&\overset{(b)}{=}\mathbb{E}\prod_{j\in\Phi_b\setminus\Phi_a} \frac{1}{1+\frac{\alpha n T p^t\|\mathbf{d}_j\|^{-\beta}}{Qp^cD}}\\
&\overset{(c)}{=} \exp\left(-2\pi \lambda \int_{r_L}^{\infty} \left(1 - \frac{1}{1+\frac{\alpha n T p^t x^{-\beta}}{Qp^cD}}\right) x \, dx\right)\\
&\overset{(d)}{=}\exp\left(-\pi \lambda \int_{r_L^2}^{\infty}  \frac{\frac{\alpha n T p^t y^{-\frac{\beta}{2}}}{Qp^cD}}{1+\frac{\alpha n T p^t y^{-\frac{\beta}{2}}}{Qp^cD}}  \, dy\right)\\
&\overset{(e)}{=} \exp\left(-\pi \lambda  H_1(Q,n,\beta,T,p_t,p_c)\right),
    \end{aligned}
\end{equation}
where \(H_1(Q,n,\beta,T,p_t,p_c)\) is defined as:
\begin{equation*}
    \frac{2}{\beta}\left(\frac{\alpha nTp^t}{QDp^c}\right)^{\frac{2}{\beta}}\left(B(\frac{2}{\beta},1-\frac{2}{\beta})-B(Q_L,\frac{2}{\beta},1-\frac{2}{\beta})\right).
\end{equation*}\(Q_L\triangleq\left(1+\frac{\alpha nTp^tr_L^{-\beta}}{DQp^c}\right)^{-1}\). In \eqref{epec_com}, \((a)\) utilizes the independence between the small-scale fading parameter and the spatial distribution of BSs. To obtain $(b)$, we harness the fact that the interference power imposed by interfering BS at the typical user follows an \(\exp(1)\). To derive $(c)$, We leverage the PGFL of the Poisson point process (PPP) in stochastic geometry. Step \((d)\) is derived by the substitution \(y=x^2\), while step \((e)\) is obtained through variable transformations and some integral strategies. \par
The expression in \eqref{epec_com} is on condition of the distances of the BSs within communication cluster, denoted by \(r_1,r_2,\cdots,r_L\). To get the final expression, we need the respective PDF of \(r_1,r_2,\cdots,r_L\). The final expectation is given by the following integral:
\begin{equation}
    \begin{aligned}
       & \int_{0}^{\infty}\cdots \int_{0}^{\infty}e^{-\pi\lambda H_1(Q,n,\beta,T,p_t,p_c)}f(r_1)\cdots f(r_L)dr_1\cdots dr_L\\ \label{final_com}
    \end{aligned}
\end{equation}
where \(f(r_L)=2\frac{\left(\lambda \pi r_L^2\right)^L}{(L-1)!r_L}\exp(-\pi\lambda r_L^2)\) is the PDF of the random variable \(r_L\), \(f(r_1)=2\pi\lambda r_1\exp(-\pi\lambda r_1^2)\) is the PDF of the random variable \(r_1\).\par
Finally, we substitute \eqref{final_com} into \eqref{p_c_1} to obtain the expression for the communication coverage probability:
\begin{equation}
\begin{aligned}
       \sum_{n=1}^{Q}(-1)^{n+1}\binom{Q}{n}\int_{0}^{\infty}\cdots \int_{0}^{\infty}e^{-\pi\lambda H_1(Q,n,\beta,T,p_t,p_c)}\\
       \times f(r_1)\cdots f(r_L)dr_1\cdots dr_L.
\end{aligned}
\end{equation}
This completes the proof.
\section{Proof of Proposition 1}
When \(L=1\), \(\beta=4\), $H_1(Q,n,\beta,T,p_t,p_c)$ simplifies as follows:
\begin{equation}
    \begin{aligned}
        &H_1(Q,n,\beta,T,p_t,p_c)\\
        &=\left(\frac{T\alpha np^t}{Qp^c}\right)^{\frac{1}{2}}r_1^2\left(B(\frac{1}{2},\frac{1}{2})-B(U_L',\frac{1}{2},\frac{1}{2})\right)\\
        &=\frac{1}{2}\left(\frac{T\alpha np^t}{Qp^c}\right)^{\frac{1}{2}} r_1^2\int_{U_L'}^{1}\frac{1}{\sqrt{t(1-t)}}dt\\
        &\overset{(a)}{=}\left(\frac{T\alpha np^t}{Qp^c}\right)^{\frac{1}{2}}\left(\frac{\pi}{2}-\arcsin{\sqrt{U_L'}}\right),\label{L=1,beta=4}
    \end{aligned}
\end{equation}
where \(U_L'\triangleq\left(1+\frac{T\alpha np^t}{Qp^c}\right)^{-1}\). Step \((a)\) is derived by evaluating the definite integral through the variable substitution \(t=\sin^2{\theta}\).\par
Substituting \eqref{L=1,beta=4} into \eqref{final_com}, the closed-form expression for the communication coverage probability under the specific case of \(L=1\), \(\beta=4\) is obtained as follows:
\begin{equation}
    \mathcal{P}_c=\sum_{n=1}^{Q}(-1)^{n+1}\binom{Q}{n}\frac{1}{1+\left(\frac{T\alpha np^t}{Qp^c}\right)^{\frac{1}{2}}\left(\frac{\pi}{2}-\arcsin(\sqrt{U_L'})\right)}.
\end{equation}
The proof has completed.
\section{Proof of LEMMA \ref{lemma_sen} }
Similar to aforementioned treatment of small-scale fading parameters, we can readily demonstrate that the small-scale fading parameters for desired echo signal power and interference signal power follow a Gamma distribution \(\Gamma(M_t-1,1)\) and an exponential distribution \(\exp\,(1)\), respectively. Building on the discussion outlined in Section \ref{sensing_performance}, we first derive the conditional expression for the Laplace transform of desired echo signal power $\mathbb{E} [e^{-zX}|r_N]$:
\begin{equation}\label{laplace_x}
\begin{aligned}
&\mathbb{E} [e^{-zX}|r_N] \\
&=\mathbb{E}\left[e^{-z\sigma^2 M_r \sum_{i=1}^N f_i\|\mathbf{d}_i\|^{-\beta}}\right] \\
&\overset{(a)}{=}\mathbb{E}_\Phi\prod_{i=1}^N\mathbb{E}_f \left[e^{-z\sigma^2 M_r f_i\|\mathbf{d}_i\|^{-\beta}}\right] \\
&\overset{(b)}{=}\mathbb{E}_\Phi\prod_{i=1}^N\left[\frac{1}{(p^s z\sigma^2 M_r \|\mathbf{d}_i\|^{-\beta}+1)^{M_t-1}}\right] \\
&\overset{(c)}{=}\exp\left({-2\pi\lambda \int_{0}^{r_N} \left(1 - \frac{1}{(p^s z \sigma^2 M_r  x^{-\beta} + 1)^{M_t - 1}}\right) xdx}\right).
\end{aligned}
\end{equation}
To derive step $(a)$, we utilize the fact that the small-scale fading parameter is independent of the spatial distribution of BSs. Step $(b)$ follows that the desired echo signal power imposed by cooperative BSs at the typical target follows a Gamma distribution $ \Gamma(M_t-1,1)$, and we employ the PGFL to obtain step $(c)$. Now we focus on the derivation of the integral in \eqref{laplace_x}:
\begin{equation}
\begin{aligned}
&\int_{0}^{r_N} \left(1 - \frac{1}{(p^s z \sigma^2 M_r x^{-\beta} + 1)^{M_t - 1}}\right) xdx\\
 &=\int_{0}^{r_N} \left( \frac{(p^s z \sigma^2 M_r x^{-\beta} + 1)^{M_t - 1}-1}{(p^s z \sigma^2 M_r x^{-\beta} + 1)^{M_t - 1}}\right) xdx\\
 &  \overset{(a)}{=}\sum_{i=1}^{Q}\binom{Q}{i}\int_{0}^{r_N} \left( \frac{(p^s z \sigma^2 M_r x^{-\beta} )^{i}}{(p^s z \sigma^2 M_r x^{-\beta} + 1)^{Q}}\right) xdx\\
  &\overset{(b)}{=}\frac{1}{\beta}H_3(M_t,z,\sigma^2,M_r,r_N,\beta,p^s),\label{ss}
\end{aligned}
\end{equation}
where \(H_3(M_t,z,\sigma^2,M_r,r_N,\beta,p^s)\) is defined as: 
\begin{equation*}
   (z\sigma^2 M_rp^s)^ {\frac{2}{\beta}} \sum_{i=1}^{Q}\binom{Q}{i}B\left(u_N,Q-i+\frac{2}{\beta},i-\frac{2}{\beta}\right).
\end{equation*}
\(u_N\triangleq(p^s z \sigma^2 M_r r_N^{-\beta} + 1)^{-1}\). Step \((a)\) is obtained using the binomial expansion, and Step \((b)\) follows from the substitution \(u=(p^s z \sigma^2 M_r  x^{-\beta} + 1)^{-1}\) and some integral strategies.\par
So the Laplace transform of the desired signal power conditioned on \(r_L\) can be given by
\begin{equation}
    \exp\left(-\frac{2\pi\lambda}{\beta} H_3(M_t,z,\sigma^2,M_r,r_N,\beta,p^s)\right).
\end{equation}
\par
To compute the Laplace transform of the interference power, it is the essential to explore the distribution of the distance between an interfering BS and the echo-receiving BS. According to the Slivnyak’s theorem in \cite{chiu2013stochastic},  for a HPPP, conditioning on the presence of a point at a specific location does not change the distribution of the rest of the process. The distance between the $N+1$-th nearest point and the nearest point can be characterized by the distance between the $N$-th nearest point and the origin. Thus, the distribution of the distance from an interfering BS to the echo-receiving BS can be equivalently treated as the distribution of the distance from a point outside \(\mathcal{O}(0,r_N)\) to the origin.\par
Based on the aforementioned analysis, the conditional expression for the Laplace transform of the interference power \(\mathbb{E} [e^{-zY}|r_1,\eta_N] \) can be given by: 
\begin{equation}
\begin{aligned}
&\mathbb{E} [e^{-zY}|r_1,\eta_N]  \\
&= \mathbb{E}\left[e^{-z\sum_{q\in \Phi_b \setminus \Phi_s} f_q \|\mathbf{d}_1 - \mathbf{d}_q\|^{-\beta} r_1^{\beta}} \right] \\
&\overset{(a)}{=}\mathbb{E}_\Phi \prod_{q\in \Phi_b \setminus \mathcal{O}(0,r_N)} \mathbb{E}_f \left[e^{-z f_q \|\mathbf{d}_1 - \mathbf{d}_q\|^{-\beta} r_1^{\beta}}\right] \\
&\overset{(b)}{=} \mathbb{E}_\Phi \prod_{q\in \Phi_b \setminus \mathcal{O}(0,r_N)} \left(\frac{1}{1 + z \|\mathbf{d}_1 - \mathbf{d}_q\|^{-\beta} r_1^{\beta}}\right) \\
&\overset{(c)}{=} \exp\left(-2\pi\lambda \int_{r_N}^{\infty} \left(1 - \frac{1}{1 + z y^{-\beta} r_1^{\beta}}\right) y \, dy \right). \label{interfen_sens}
\end{aligned}
\end{equation}
Step \((a)\) is derived using the independence between the small-scale fading parameter and the BSs positions, step \((b)\) is based on the statistical property that $f_q$ follows an exponential distribution $\exp(1)$, and step \((c)\) is derived using the PFGL of the HPPP. To evaluate the integral item in \eqref{interfen_sens}, we perform a variable substitution to obtain a tractable expression:
\begin{equation}
    \begin{aligned}
&\int_{r_N}^{\infty} \left(1 - \left(\frac{1}{1+zy^{-\beta}r_1^{\beta}}\right) ydy\right)\\
&\overset{(a)}{=}r_1^2\int_{\eta_N}^{1} \left(\frac{1}{1+zt^{\beta}}t^{-3}dt\right)\\
&\overset{(b)}{=} r_1^2H_4(z,\beta,\eta_N),
\end{aligned}
\end{equation}
where \(H_4(z,\beta,\eta_N)\) is defined as
\begin{equation*}
    \frac{z^{\frac{2}{\beta}}}{\beta}\left(B\left(\frac{2}{\beta},1-\frac{2}{\beta}\right)-B\left(v_N,\frac{2}{\beta},1-\frac{2}{\beta}\right)\right).
\end{equation*}
\(v_N\triangleq\frac{1}{1+z\eta_L^{\beta}}\), the distance ratio \(\eta_N=\frac{r_1}{r_N}\). Step \((a)\) is obtained by the substitution \(t=\frac{r_1}{y}\), and step \((b)\) follows from the definition \(v=(1+zt^{\beta})^{-1}\). Therefore, the Laplace transform of the interference signal power conditioned on the \(r_1,\eta_N\) is given by
\begin{equation}
   \begin{aligned}
    \exp\left(-2\pi\lambda r_1^2 H_4(z,\beta,\eta_N)\right).
\end{aligned} 
\end{equation}
This completes the proof.
\section{Proof of Theorem 2}
Considering that the results in Lemma \ref{lemma_sen} is conditioned on the distances to the farthest BS, the nearest BS, and their distance ratio, the PDFs of these random variables are required. By integrating over these distributions, the final expression for the radar information rate can be obtained.\par
Firstly, the expression of the Laplace transform of the desired echo signal power is conditioned on \(r_N=\|\mathbf{d}_N\|\),the distance to the farthest BS. Therefore, the final expression can be obtained by integrating the product of the conditional expression and the PDF of \(r_N\).
\begin{equation}
   \int_0^{\infty} e^{-\frac{2\pi\lambda}{\beta} H_3(M_t,z,\sigma^2,M_r,r_N,\beta,p^s)}  f(r_N)  \, dr_N,\label{eq:24}
\end{equation}
where \(f(r_N)=\frac{2(\lambda \pi r_N^2)^{N}}{\Gamma(N)r_N}e^{-\lambda \pi r_N^2}\).\par
Secondly, the expression of the Laplace transform of the interference signal power conditioned on \(r_1=\|\mathbf{d}_1\|,\eta_{N}=\frac{r_1}{r_N}\). The final expression can be obtained by integrating the product of the conditional expression, the PDF of \(\eta_N\), and the PDF of \(r_1\).
\begin{equation}
    \int_{0}^{1} \int_{0}^{\infty}\exp(-2\pi\lambda r_1^2H_4(z,\beta,\eta_N))f(r_1) f(\eta_N)\, dr_1 \,d\eta_N,\label{eq:15}
\end{equation}
where \(f(r_1)=2\lambda r_1 e^{-\lambda \pi r_1^2}\) and \(f(\eta_N)=2(N-1)\eta_N(1-\eta_N^2)^{N-2}\). To simplify the expression in \eqref{eq:15}, we process this integral as follows
\begin{equation}
    \begin{aligned}
 & \int_{0}^{1} \int_{0}^{\infty}\exp(-2\pi\lambda r_1^2 H_4(z,\beta,\eta_L))f(r_1) f(\eta_N) dr_1 d\eta_N\\ 
 & \overset{(a)}{=}\int_{0}^{1}\int_{0}^{\infty} e^{-\pi\lambda r_1^2 \left(1+2 H_4(z,\beta,\eta_L)\right)}2\lambda r_1  f(\eta_N) dr_1 d\eta_N\\ 
 &\overset{(b)}{=}\int_{0}^{1}\frac{1}{1 + 2H_4(z,\beta,\eta_N)} f(\eta_N) \, d\eta_N \label{eq:25}
\end{aligned}
\end{equation}
Step \((a)\) is derived from the expression for the PDF of \(r_1\), while Step \((b)\) results from the integration with respect to \(r_1\).\par
Finally, by substituting \eqref{eq:24} and \eqref{eq:25} into \eqref{eq:rate}, the expression for the radar information rate is obtained.
The proof has completed.
\section{Proof of Proposition 2}
when $N=1$, the expression for the sensing SIR can be simplified as follows: 
\begin{equation}
    SIR_s=\sigma^2 M_r\frac{\|\mathbf{d}_1\|^{-2\beta}p^s f_1}{\sum_{q\in \Phi_b \setminus \Phi_s}p^t f_q\|\mathbf{d}_1-\mathbf{d}_q\|^{-\beta}}.
\end{equation}
Accordingly, the radar information rate can be expressed as follows:
\begin{equation}
    \begin{aligned}
        R_s=&\mathbb{E}[\log(1+SIR_s)]\\
        =&\int_{0}^{\infty} \frac{1-\mathbb{E}[e^{-zX}]}{z}\mathbb{E}[e^{-zY}]\, dz,\\
        \label{pos}
    \end{aligned}
\end{equation}
where \(X\) $\triangleq$ \(\sigma^2 M_r p^s f_1\), and \(Y\) $\triangleq$ \(\|\mathbf{d}_1\|^{2\beta}\sum_{q\in \Phi_b \setminus \Phi_s}p^t f_q\|\mathbf{d}_1-\mathbf{d}_q\|^{-\beta}\). Likewise, we condition the analysis on the distance between the target and the echo-receiving BS, i.e., \(\|\mathbf{d}_1\|=r_1\). We first focus on the derivation on the Laplace transform with respect to the desired signal power:
\begin{equation}
    \begin{aligned}
        \mathbb{E}[e^{-zX}]&= \mathbb{E}[{e^{-z\sigma^2M_r p^s f_1}}]\\
        &\overset{(a)}{=}\frac{1}{(1+z\sigma^2M_rp^s)^{M_t-1}}.
        \label{part_1}
    \end{aligned}
\end{equation}
To obtain the relation \((a)\), we utilize the property of the Gamma distributed random variable \(f_1 \sim\Gamma(M_t-1,1)\).\par
As shown Fig.\ref{fig:interference hole}, the nearest BS is the echo-receiving BS, and the distance between it and the target is denoted by \(r_1\). It is evident that no BS exists within the circular region centered at the target with radius \(r_1\). Here, we use \(x\) to denote the distance from an interference BS to the nearest sensing BS. The distribution range of interference hole in the angular domain is approximately \([-\phi(x),\phi(x)]\) where \(\phi(x)=\arccos{\frac{x}{2r_1}}\) is derived from geometric relations. In the radial domain, the interference hole spans the range spans \([0,2r_1]\). In the calculation, we can first ignore the impact of interference hole and then eliminate the effect of interference BS distributions within the interference hole from a polar coordinate perspective. The detailed derivation of this process is presented below.\par
\begin{equation}
    \begin{aligned}
       & \mathbb{E}[e^{-zY}|r_1] \\
        &= \mathbb{E}\left[e^{-zr_1^{2\beta}\sum_{q\in \Phi_b \setminus \Phi_s} p^t f_q\|\mathbf{d}_1-\mathbf{d}_q\|^{-\beta}}\right] \\
        &= \mathbb{E}\left[\prod_{q\in \Phi_b \setminus \mathcal{O}(0,r_1)} 
            \mathbb{E}_{f_q}\left[e^{-zr_1^{2\beta}p^t f_q\|\mathbf{d}_1-\mathbf{d}_q\|^{-\beta}}\right] 
        \right] \\
        &\overset{(a)}{=} \mathbb{E}\left[\prod_{q\in \Phi_b \setminus \mathcal{O}(0,r_1)} 
            \frac{1}{1+zr_1^{2\beta}p^t \|\mathbf{d}_1-\mathbf{d}_q\|^{-\beta}} 
        \right] \\
        &\overset{(b)}{=} \exp\left(-2\pi \lambda \int_{0}^{\infty} 
            \left(1 - \frac{1}{1+zr_1^{2\beta}p^t  x^{-\beta}}\right) x \,dx \right) \\
        &\phantom{=} \times \exp\left(\lambda \int_{0}^{2r_1} 2\arccos{\frac{x}{2r_1}} 
            \left(1 - \frac{1}{1+zr_1^{-2\beta}p^t  x^{-\beta}}\right) x \,dx \right) \\
        &\overset{(c)}{=} \exp\left(-\pi\lambda r_1^{4} \frac{2}{\beta} (zp^t)^{\frac{2}{\beta}} 
            B\left(\frac{2}{\beta}, 1-\frac{2}{\beta}\right)\right)  \\
        &\phantom{=} \times \exp\left(\lambda r_1^2 \int_{0}^{2} 2\arccos{\frac{t}{2}} 
            \frac{zp^t   r_1^{\beta} t^{-\beta}}{1+zp^t r_1^{\beta}  t^{-\beta}} t\, dt\right).
    \end{aligned}
\end{equation}
We utilize the property of the exponential random variables \(f_q\sim \exp\,(1)\) to derive relation \((a)\), and apply the PGFL of a HPPP to obtain relation \((b)\). The first term in relation \((b)\) accounts for the impact of entire interfering BSs distributed over the region, covering the full angular domain \(0\) to \(2\pi\) and the radial domain \(0\) to \(\infty\) in polar coordinates. The second term captures the effect of the interference hole region, which spans the angular rangee \(-\phi(x)\) to \(\phi(x)\) and the radial range \(0\) to \(2r_1\). Finally, relation \((c)\) is obtained by applying integration techniques with the substitution \(t=\frac{x}{r_1}\). By integrating over the distribution of the distance \(r_1\), we obtain:
\begin{equation}
    \begin{aligned}
         \mathbb{E}[e^{-zY}]=&\int_{0}^{\infty} f(r_1)\exp\left(-\pi\lambda r_1^{4}\frac{2}{\beta} (zp^t)^{\frac{2}{\beta}} 
            B\left(\frac{2}{\beta}, 1-\frac{2}{\beta}\right)\right) \\
             &\phantom{=}\times \exp\left(\lambda r_1^2 \int_{0}^{2} 2\arccos{\frac{t}{2}} 
            \frac{zp^t   r_1^{\beta} t^{-\beta}}{1+zp^t   r_1^{\beta} t^{-\beta}} t\, dt\right),  \label{part_2}
    \end{aligned}
\end{equation}
By substituting \eqref{part_1} and \eqref{part_2} into \eqref{pos}, the radar information rate under the specified conditions can be expressed as:
\begin{equation}
    \begin{aligned}
        R_s=&\int_{0}^{\infty}\frac{1}{z}\left(1-\frac{1}{(1+z\sigma^2M_rp^s)^{M_t-1}}\right)
        \\ &\phantom{=}\int_{0}^{\infty} f(r_1)\exp\left(-\pi\lambda r_1^{4} \frac{2}{\beta} (zp^t)^{\frac{2}{\beta}} 
            B\left(\frac{2}{\beta}, 1-\frac{2}{\beta}\right)\right) \\
             &\phantom{=}\times \exp\left(\lambda r_1^2 \int_{0}^{2} 2\arccos{\frac{t}{2}} 
            \frac{zp^t  r_1^{\beta} t^{-\beta}}{1+zp^t  r_1^{\beta} t^{-\beta}} t\, dt\right)\,dz.
    \end{aligned}
\end{equation}
This completes the proof.
\ifCLASSOPTIONcaptionsoff
  \newpage
\fi

 \bibliographystyle{IEEEtran}
 \bibliography{ref}

\begin{thebibliography}{10}
\providecommand{\url}[1]{#1}
\csname url@samestyle\endcsname
\providecommand{\newblock}{\relax}
\providecommand{\bibinfo}[2]{#2}
\providecommand{\BIBentrySTDinterwordspacing}{\spaceskip=0pt\relax}
\providecommand{\BIBentryALTinterwordstretchfactor}{4}
\providecommand{\BIBentryALTinterwordspacing}{\spaceskip=\fontdimen2\font plus
\BIBentryALTinterwordstretchfactor\fontdimen3\font minus
  \fontdimen4\font\relax}
\providecommand{\BIBforeignlanguage}[2]{{%
\expandafter\ifx\csname l@#1\endcsname\relax
\typeout{** WARNING: IEEEtran.bst: No hyphenation pattern has been}%
\typeout{** loaded for the language `#1'. Using the pattern for}%
\typeout{** the default language instead.}%
\else
\language=\csname l@#1\endcsname
\fi
#2}}
\providecommand{\BIBdecl}{\relax}
\BIBdecl

\bibitem{8869705}
W.~Saad, M.~Bennis, and M.~Chen, ``A vision of {6G} wireless systems:
  Applications, trends, technologies, and open research problems,'' \emph{IEEE
  Network}, vol.~34, no.~3, pp. 134--142, May 2020.

\bibitem{9705498}
A.~Liu, Z.~Huang, M.~Li, Y.~Wan, W.~Li, T.~X. Han, C.~Liu, R.~Du, D.~K.~P. Tan,
  J.~Lu, Y.~Shen, F.~Colone, and K.~Chetty, ``A survey on fundamental limits of
  integrated sensing and communication,'' \emph{IEEE Commun. Surv. Tutorials},
  vol.~24, no.~2, pp. 994--1034, Secondquarter 2022.

\bibitem{9945983}
F.~Dong, F.~Liu, Y.~Cui, W.~Wang, K.~Han, and Z.~Wang, ``Sensing as a service
  in 6g perceptive networks: A unified framework for {ISAC} resource
  allocation,'' \emph{IEEE Trans. Wireless Commun.}, vol.~22, no.~5, pp.
  3522--3536, May 2023.

\bibitem{9557830}
W.~Yuan, Z.~Wei, S.~Li, J.~Yuan, and D.~W.~K. Ng, ``Integrated sensing and
  communication-assisted orthogonal time frequency space transmission for
  vehicular networks,'' \emph{IEEE J. Sel. Top. Signal Process.}, vol.~15,
  no.~6, pp. 1515--1528, Nov 2021.

\bibitem{9830717}
X.~Cheng, D.~Duan, S.~Gao, and L.~Yang, ``Integrated sensing and communications
  {(ISAC)} for vehicular communication networks {(VCN)},'' \emph{IEEE Internet
  Things J.}, vol.~9, no.~23, pp. 23\,441--23\,451, Dec 2022.

\bibitem{9755276}
Z.~Wei, F.~Liu, C.~Masouros, N.~Su, and A.~P. Petropulu, ``Toward
  multi-functional {6G} wireless networks: Integrating sensing, communication,
  and security,'' \emph{IEEE Commun. Mag.}, vol.~60, no.~4, pp. 65--71, April
  2022.

\bibitem{7465731}
E.~Cianca, M.~De~Sanctis, and S.~Di~Domenico, ``Radios as sensors,'' \emph{IEEE
  Internet Things J.}, vol.~4, no.~2, pp. 363--373, April 2017.

\bibitem{10566041}
Y.~Liu, X.~Liu, Z.~Liu, Y.~Yu, M.~Jia, Z.~Na, and T.~S. Durrani, ``Secure rate
  maximization for {ISAC-UAV} assisted communication amidst multiple
  eavesdroppers,'' \emph{IEEE Trans. Veh. Technol.}, vol.~73, no.~10, pp.
  15\,843--15\,847, Oct 2024.

\bibitem{10054402}
Q.~Zhu, M.~Li, R.~Liu, and Q.~Liu, ``Joint transceiver beamforming and
  reflecting design for active {RIS}-aided {ISAC} systems,'' \emph{IEEE Trans.
  Veh. Technol.}, vol.~72, no.~7, pp. 9636--9640, July 2023.

\bibitem{10527368}
Z.~Yu, X.~Hu, C.~Liu, and M.~Peng, ``{IRS}-aided non-orthogonal {ISAC} systems:
  Performance analysis and beamforming design,'' \emph{IEEE Trans. Green
  Commun. Networking}, vol.~8, no.~4, pp. 1930--1942, Dec 2024.

\bibitem{10644091}
E.~Memisoglu, M.~B. Janjua, and H.~Arslan, ``Power-efficient time-domain
  scheduling for {ISAC} beamforming,'' \emph{IEEE Wireless Commun. Lett.},
  vol.~13, no.~10, pp. 2837--2841, Oct 2024.

\bibitem{9585321}
J.~A. Zhang, M.~L. Rahman, K.~Wu, X.~Huang, Y.~J. Guo, S.~Chen, and J.~Yuan,
  ``Enabling joint communication and radar sensing in mobile networks—a
  survey,'' \emph{IEEE Commun. Surv. Tutorials}, vol.~24, no.~1, pp. 306--345,
  Firstquarter 2022.

\bibitem{10012421}
Z.~Wei, H.~Qu, Y.~Wang, X.~Yuan, H.~Wu, Y.~Du, K.~Han, N.~Zhang, and Z.~Feng,
  ``Integrated sensing and communication signals toward {5G-A} and {6G}: A
  survey,'' \emph{IEEE Internet Things J.}, vol.~10, no.~13, pp.
  11\,068--11\,092, July 2023.

\bibitem{9540344}
J.~A. Zhang, F.~Liu, C.~Masouros, R.~W. Heath, Z.~Feng, L.~Zheng, and
  A.~Petropulu, ``An overview of signal processing techniques for joint
  communication and radar sensing,'' \emph{IEEE J. Sel. Top. Signal Process.},
  vol.~15, no.~6, pp. 1295--1315, Nov 2021.

\bibitem{10159012}
Z.~Liu, S.~Aditya, H.~Li, and B.~Clerckx, ``Joint transmit and receive
  beamforming design in full-duplex integrated sensing and communications,''
  \emph{IEEE J. Sel. Areas Commun.}, vol.~41, no.~9, pp. 2907--2919, Sep. 2023.

\bibitem{9860459}
W.~Jiang, Z.~Wei, and Z.~Feng, ``Toward multiple integrated sensing and
  communication base station systems: Collaborative precoding design with power
  constraint,'' in \emph{2022 IEEE 95th Vehicular Technology Conference:
  (VTC2022-Spring)}, June 2022, pp. 1--5.

\bibitem{10892128}
X.~Li, Q.~Zhu, Y.~Chen, and Y.~Yuan, ``Distributed multi-node cooperative
  integrated sensing and communication systems: Joint beamforming and grouping
  design,'' \emph{IEEE Internet Things J.}, pp. 1--1, 2025.

\bibitem{10540489}
C.~Dou, N.~Huang, Y.~Wu, L.~Qian, Z.~Shi, and T.~Q.~S. Quek, ``Integrated
  sensing and communication enabled multidevice multitarget cooperative
  sensing: A fairness-aware design,'' \emph{IEEE Internet Things J.}, vol.~11,
  no.~17, pp. 29\,190--29\,201, Sep. 2024.

\bibitem{10833779}
K.~Meng, C.~Masouros, K.-K. Wong, A.~P. Petropulu, and L.~Hanzo, ``Integrated
  sensing and communication meets smart propagation engineering: Opportunities
  and challenges,'' \emph{IEEE Network}, vol.~39, no.~2, pp. 278--285, March
  2025.

\bibitem{ren2024optimal}
Z.~Ren, Y.~Yu, H.~Ren, C.~Pan, and J.~Wang, ``What is the optimal inter-site
  distance in multi-{BS} cooperative sensing?'' \emph{Sci. China Inf. Sci.},
  vol.~67, no.~12, p. 229302, 2024.

\bibitem{10726912}
K.~Meng, C.~Masouros, A.~P. Petropulu, and L.~Hanzo, ``Cooperative {ISAC}
  networks: Opportunities and challenges,'' \emph{IEEE Wireless Commun.}, pp.
  1--8, 2024.

\bibitem{10680299}
A.~Khalili, A.~Rezaei, D.~Xu, F.~Dressler, and R.~Schober, ``Efficient uav
  hovering, resource allocation, and trajectory design for isac with limited
  backhaul capacity,'' \emph{IEEE Trans. Wireless Commun.}, vol.~23, no.~11,
  pp. 17\,635--17\,650, Nov 2024.

\bibitem{6287527}
H.-S. Jo, Y.~J. Sang, P.~Xia, and J.~G. Andrews, ``Heterogeneous cellular
  networks with flexible cell association: A comprehensive downlink sinr
  analysis,'' \emph{IEEE Trans. Wireless Commun.}, vol.~11, no.~10, pp.
  3484--3495, October 2012.

\bibitem{6042301}
J.~G. Andrews, F.~Baccelli, and R.~K. Ganti, ``A tractable approach to coverage
  and rate in cellular networks,'' \emph{IEEE Trans. Commun.}, vol.~59, no.~11,
  pp. 3122--3134, November 2011.

\bibitem{6658810}
H.~S. Dhillon and J.~G. Andrews, ``Downlink rate distribution in heterogeneous
  cellular networks under generalized cell selection,'' \emph{IEEE Wireless
  Commun. Lett.}, vol.~3, no.~1, pp. 42--45, February 2014.

\bibitem{6596082}
H.~S. Dhillon, M.~Kountouris, and J.~G. Andrews, ``Downlink {MIMO} hetnets:
  Modeling, ordering results and performance analysis,'' \emph{IEEE Trans.
  Wireless Commun.}, vol.~12, no.~10, pp. 5208--5222, October 2013.

\bibitem{6881662}
A.~K. Gupta, H.~S. Dhillon, S.~Vishwanath, and J.~G. Andrews, ``Downlink
  multi-antenna heterogeneous cellular network with load balancing,''
  \emph{IEEE Trans. Commun.}, vol.~62, no.~11, pp. 4052--4067, Nov 2014.

\bibitem{9542942}
J.~He, Y.~J. Chun, and H.~C. So, ``A unified analytical framework for
  {RSS-Based} localization systems,'' \emph{IEEE Internet Things J.}, vol.~9,
  no.~9, pp. 6506--6519, May 2022.

\bibitem{8587144}
J.~D. Roth, M.~Tummala, and J.~C. McEachen, ``Fundamental implications for
  location accuracy in ultra-dense {5G} cellular networks,'' \emph{IEEE Trans.
  Veh. Technol.}, vol.~68, no.~2, pp. 1784--1795, Feb 2019.

\bibitem{9580712}
S.~S. Ram, G.~Singh, and G.~Ghatak, ``Optimization of radar parameters for
  maximum detection probability under generalized discrete clutter conditions
  using stochastic geometry,'' \emph{IEEE Open J. Signal Process.}, vol.~2, pp.
  571--585, 2021.

\bibitem{7412750}
J.~Schloemann, H.~S. Dhillon, and R.~M. Buehrer, ``A tractable analysis of the
  improvement in unique localizability through collaboration,'' \emph{IEEE
  Trans. Wireless Commun.}, vol.~15, no.~6, pp. 3934--3948, June 2016.

\bibitem{10556618}
X.~Gan, C.~Huang, Z.~Yang, X.~Chen, J.~He, Z.~Zhang, C.~Yuen, Y.~Liang~Guan,
  and M.~Debbah, ``Coverage and rate analysis for integrated sensing and
  communication networks,'' \emph{IEEE J. Sel. Areas Commun.}, vol.~42, no.~9,
  pp. 2213--2227, Sep. 2024.

\bibitem{10769538}
K.~Meng, C.~Masouros, A.~P. Petropulu, and L.~Hanzo, ``Cooperative {ISAC}
  networks: Performance analysis, scaling laws, and optimization,'' \emph{IEEE
  Trans. Wireless Commun.}, vol.~24, no.~2, pp. 877--892, Feb 2025.

\bibitem{10735119}
K.~Meng, C.~Masouros, G.~Chen, and F.~Liu, ``Network-level integrated sensing
  and communication: Interference management and {BS} coordination using
  stochastic geometry,'' \emph{IEEE Trans. Wireless Commun.}, vol.~23, no.~12,
  pp. 19\,365--19\,381, Dec 2024.

\bibitem{rir}
Y.~Yang and R.~S. Blum, ``{MIMO} radar waveform design based on mutual
  information and minimum mean-square error estimation,'' \emph{IEEE Trans.
  Aerosp. Electron. Syst.}, vol.~43, no.~1, pp. 330--343, January 2007.

\bibitem{rir2}
B.~Tang and J.~Li, ``Spectrally constrained {MIMO} radar waveform design based
  on mutual information,'' \emph{IEEE Trans. Signal Process.}, vol.~67, no.~3,
  pp. 821--834, Feb 2019.

\bibitem{10695929}
Y.~Jiang, F.~Meng, X.~Li, X.~Li, G.~Zhu, K.~Han, and Q.~Shi, ``Coverage
  analysis for air-ground integrated-sensing-and-communication networks,'' in
  \emph{2024 International Conference on Ubiquitous Communication (Ucom)}, July
  2024, pp. 455--459.

\bibitem{10683162}
X.~Gan, C.~Huang, Z.~Yang, X.~Chen, J.~He, Z.~Zhang, C.~Yuen, Y.~L. Guan, and
  M.~Debbah, ``Toward a unified analytical framework for {ISAC} fundamentals in
  cellular networks,'' in \emph{2024 IEEE 99th Vehicular Technology Conference
  (VTC2024-Spring)}, June 2024, pp. 1--6.

\bibitem{6856159}
X.~Zhang and M.~Haenggi, ``A stochastic geometry analysis of inter-cell
  interference coordination and intra-cell diversity,'' \emph{IEEE Trans.
  Wireless Commun.}, vol.~13, no.~12, pp. 6655--6669, Dec 2014.

\bibitem{6376184}
K.~Huang and J.~G. Andrews, ``An analytical framework for multicell cooperation
  via stochastic geometry and large deviations,'' \emph{IEEE Trans. Inf.
  Theory}, vol.~59, no.~4, pp. 2501--2516, April 2013.

\bibitem{chiu2013stochastic}
S.~N. Chiu, D.~Stoyan, W.~S. Kendall, and J.~Mecke, \emph{Stochastic geometry
  and its applications}.\hskip 1em plus 0.5em minus 0.4em\relax John Wiley \&
  Sons, 2013.

\bibitem{goldsmith2005wireless}
A.~Goldsmith, \emph{Wireless communications}.\hskip 1em plus 0.5em minus
  0.4em\relax Cambridge university press, 2005.

\bibitem{10736660}
Z.~Zhang, H.~Ren, C.~Pan, S.~Hong, D.~Wang, J.~Wang, and X.~You, ``Target
  localization in cooperative isac systems: A scheme based on 5g nr ofdm
  signals,'' \emph{IEEE Trans. Commun.}, pp. 1--1, 2024.

\bibitem{6932503}
T.~Bai and R.~W. Heath, ``Coverage and rate analysis for {Millimeter-Wave}
  cellular networks,'' \emph{IEEE Trans. Wireless Commun.}, vol.~14, no.~2, pp.
  1100--1114, Feb 2015.

\bibitem{5407601}
K.~A. Hamdi, ``A useful lemma for capacity analysis of fading interference
  channels,'' \emph{IEEE Trans. Commun.}, vol.~58, no.~2, pp. 411--416,
  February 2010.

\end{thebibliography}
\vspace{12pt}
\end{document}